\documentclass[twocolumn, journal]{IEEEtran}
\IEEEoverridecommandlockouts
\usepackage{color}
\usepackage{graphicx}
\usepackage{epstopdf}
\usepackage{amsmath}
\usepackage{amssymb}
\usepackage{algorithm}
\usepackage{algorithmic}
\usepackage{amsmath}
\usepackage{multirow}
\usepackage{booktabs}
\usepackage{array}
\usepackage{amsthm}
\usepackage{stfloats}
\usepackage{caption}
\usepackage{subfigure}
\usepackage{bm}
\usepackage{setspace}
\usepackage{diagbox}
\usepackage{flushend}
\usepackage{enumitem}
\usepackage{comment}

\newcommand{\be}{\begin{equation}}
\newcommand{\ee}{\end{equation}}
\newcommand{\bea}{\begin{eqnarray}}
\newcommand{\eea}{\end{eqnarray}}
\newcommand{\ba}{\begin{array}}
\newcommand{\ea}{\end{array}}

\allowdisplaybreaks[4]

\def\BibTeX{{\rm B\kern-.05em{\sc i\kern-.025em b}\kern-.08em
    T\kern-.1667em\lower.7ex\hbox{E}\kern-.125emX}}
\begin{document}

\title{Clutter Suppression in ISAC Systems with Compound Reconfigurable Antenna Arrays
\thanks{M. Liu and M. Li are with the School of Information and Communication Engineering, Dalian University of Technology, Dalian 116024, China (e-mail: liumengzhen@mail.dlut.edu.cn, mli@dlut.edu.cn).}
\thanks{R. Liu and A. Lee Swindlehurst are with the Nhu Department of Electrical Engineering and Computer Science, University of California, Irvine, CA 92697 (e-mail: rangl2@uci.edu; swindle@uci.edu).}
\thanks{Q. Liu is with the School of Computer Science and Technology, Dalian University of Technology, Dalian 116024, China (e-mail: qianliu@dlut.edu.cn).}}

\author{Mengzhen Liu,~\IEEEmembership{Graduate Student Member,~IEEE,}
        Ming Li,~\IEEEmembership{Senior Member,~IEEE,}
        Rang Liu,~\IEEEmembership{Member,~IEEE,} \\ Qian Liu,~\IEEEmembership{Member,~IEEE,} and A. Lee Swindlehurst,~\IEEEmembership{Life Fellow,~IEEE}
\vspace{-0.0 cm}
}

\maketitle
\pagestyle{empty}  
\thispagestyle{empty} 

\begin{abstract}
Integrated sensing and communication (ISAC) systems often suffer severe performance degradation due to strong clutter echoes, and spatial-only beamforming is often inadequate for realistic array sizes. This paper addresses clutter suppression in ISAC by leveraging compound reconfigurable antenna (CRA) arrays, which simultaneously enable dynamic adjustment of both radiation patterns and polarization states, thus substantially expanding the degrees of freedom available in the electromagnetic (EM) domain. We develop a unified compound channel model that integrates virtual angular-domain responses, spatial propagation, and polarization rotation/depolarization. Leveraging statistical information about target and clutter covariances, we formulate a joint EM-domain and baseband-domain optimization aimed at maximizing the radar signal-to-clutter-plus-noise ratio (SCNR). The formulation also enforces multiuser downlink signal-to-interference-plus-noise ratio constraints, a total transmit-power budget, and finite-codebook EM-mode selection. The resulting nonconvex mixed-integer problem is tackled by an alternating algorithm that combines fractional programming and majorization-minimization with second-order cone programming-based updates and a penalty relaxation for mode selection.  Extensive simulations in QuaDRiGa-based channel environments validate the effectiveness of the proposed CRA array design, demonstrating up to 11 dB SCNR improvements over conventional beamforming methods relying solely on baseband-domain optimization and confirming the substantial benefits of fully exploiting EM-domain reconfigurability for clutter-rich ISAC scenarios.

\end{abstract}


\section{Introduction}
Integrated sensing and communication (ISAC) has become a potential enabling technology for future wireless networks, especially in the context of sixth-generation (6G) systems \cite{F. Liu 2022 ISAC}-\cite{Fan Liu SPM 2023}. By integrating radar sensing and wireless communications on shared hardware and spectrum resources, ISAC reduces redundancy, enhances spectral efficiency, and supports various critical applications such as autonomous driving, smart manufacturing, and environmental monitoring. Nevertheless, a significant practical challenge in ISAC implementations is the presence of clutter, defined as unwanted echoes or interference from surrounding objects, which can severely degrade target detection and parameter estimation performance \cite{Zhang-JSTSP-2021}-\cite{Luo-TWC-2024}. The clutter issue is particularly pronounced in dense urban environments, where multiple transmitters and receivers coexist among complex static structures like buildings and mobile scatterers such as vehicles and pedestrians, resulting in rich and dynamic clutter.

In multi-antenna ISAC systems, clutter mitigation is often pursued through spatial-domain signal processing, where beamforming and precoding exploit spatial degrees of freedom (DoFs) to enhance the radar signal-to-clutter-plus-noise ratio (SCNR) by steering energy toward the target and placing nulls toward dominant clutter scatterers \cite{Li-TAES-2017}, \cite{Liao-TCOM-2023}. Beyond conventional array processing, several complementary approaches have been explored, such as reconfigurable intelligent surfaces (RIS) for environment shaping \cite{RLiu-JSTSP-2022,FWang-RadarConf-2023}, space-time adaptive processing (STAP) for joint spatial–temporal suppression \cite{RLiu-JSAC-2022}, \cite{Zhang-TCCN-2025}, and cell-free architectures for distributed spatial diversity \cite{SRivetti-arxiv-2025}, \cite{Jiang-arxiv-2025}.

However, the effectiveness of these spatial-domain enhancements remains fundamentally limited by practical constraints on antenna array design, including the number of antenna elements, RF chains, and physical array aperture. In realistic 6G deployments, increasing the array size or antenna count is often restricted by cost considerations, power consumption, form factor constraints, and deployment regulations. Consequently, the incremental benefit of adding additional antenna elements may diminish under these practical limitations. As a result, relying solely on spatial-domain techniques may not suffice to meet the demanding sensing performance requirements encountered in clutter-rich scenarios, thus motivating the exploration of additional DoFs beyond the spatial domain.

Recent advances in reconfigurable antenna (RA) technologies introduce promising complementary electromagnetic (EM)-domain DoFs \cite{RA book}-\cite{H. Li RA}. By dynamically adjusting antenna radiation characteristics, including radiation patterns \cite{K.K.Wong R. Murch 2024 Antenna coding pixel,Z. Gao 2024 EM domain}, polarization states \cite{R. W. Heath 2024 polarization,R. Liu security RA}, and operating frequencies \cite{R. Murch 2024 DUAL,K. Chen 2025 DUAL}, EM-domain reconfigurability can provide new mechanisms for interference/clutter suppression that are largely independent of baseband (BB) digital beamforming. In particular, pattern reconfigurability can reshape mainlobe/sidelobe shapes to strengthen spatial selectivity, while reconfiguring the polarization can exploit polarization-dependent scattering and depolarization differences between targets and clutter scatterers to improve target discrimination. Despite these benefits, most existing studies focus on pattern or polarization reconfigurability alone,  which limits the gains achievable when multiple EM-domain DoFs are simultaneously available \cite{R. Liu WCM 2025,Mengzhen CM 2025}.

\begin{figure}[!t]
    \centering
    \vspace{-0.0cm}
    \subfigure{{\includegraphics[width=1.7 in]{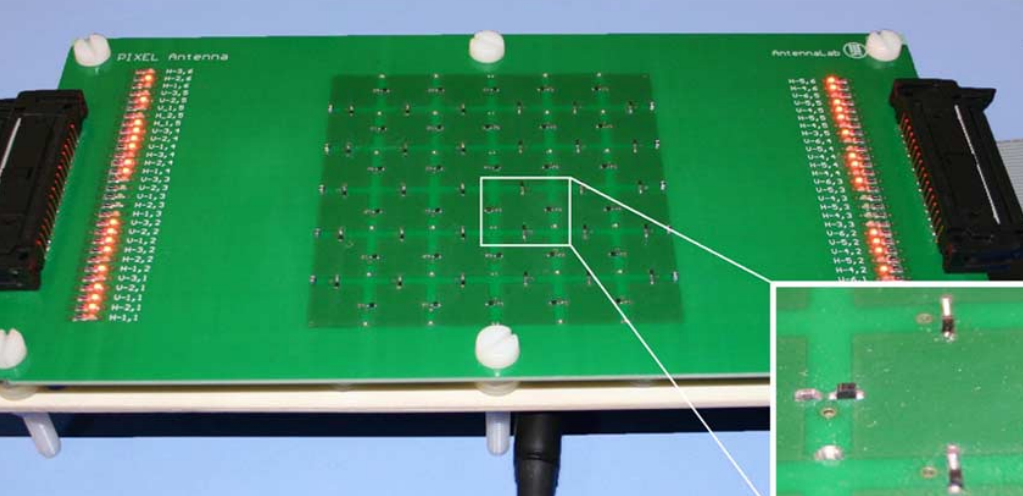}}}
    \subfigure{{\includegraphics[width=1.7 in]{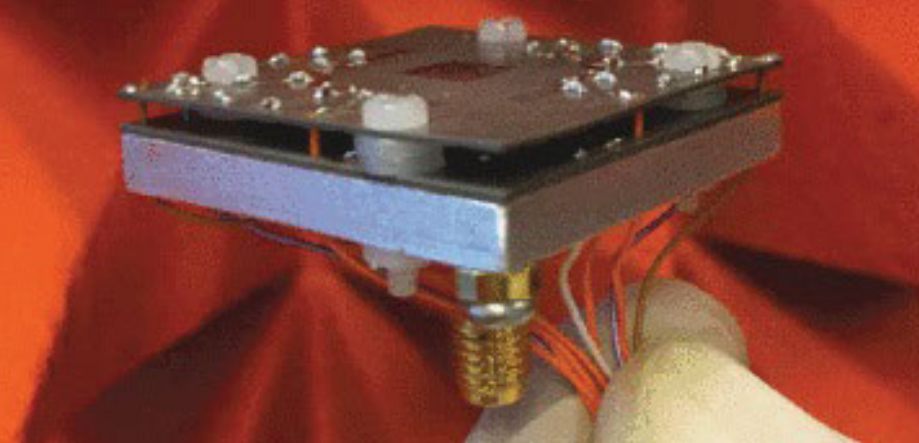}}}
    \caption{Hardware CRA prototypes; Left: Pixel-based antenna \cite{CRA hardware pixel}; Right: Cavity-backed reconfigurable microstrip antenna  \cite{CRA hardware Cavity}.}

    \label{fig:hardware}
\end{figure}

To overcome this limitation, compound reconfigurable antennas (CRAs) that integrate both radiation-pattern and polarization reconfigurability within a single antenna element have been proposed \cite{Review CRA,CRA pattern polarization}. Representative CRA prototypes, such as the pixel-based antennas and cavity-backed reconfigurable microstrip antennas shown in Fig.~\ref{fig:hardware} can realize diverse radiation patterns and polarization states through internal reconfiguration mechanisms. While a single CRA element provides joint pattern and polarization agility, deploying CRAs in an array enables element-wise EM-mode selection together with coherent multi-antenna transmission and reception. Consequently, by jointly optimizing the EM-domain configuration and the BB-domain beamforming, CRA arrays can significantly enhance clutter suppression and radar sensing accuracy without compromising the communication quality of service (QoS).

Nevertheless, several critical issues remain open: \textit{i}) Developing a unified and analytically tractable channel model that jointly captures virtual angular-domain structure, spatial propagation, and polarization rotation/depolarization effects; \textit{ii}) Designing computationally efficient joint EM- and BB-domain transceivers under practical multiuser signal-to-interference-plus-noise ratio (SINR)/QoS constraints with discrete EM-mode selection; \textit{iii}) Quantitatively characterizing the respective gains of pattern versus polarization reconfigurability, as well as the performance-complexity tradeoff induced by the CRA codebook resolution.

In this paper, we address the above challenges by proposing a comprehensive joint EM- and BB-domain design framework for CRA-array-empowered ISAC in clutter-rich environments. The key contributions of this paper are summarized below.
\begin{itemize}
\item \textbf{Unified Compound EM-Domain Modeling:} We propose a comprehensive mathematical model for EM-domain precoding and combining using CRA arrays. The model integrates three critical components: Virtual angular channels (discrete spatial direction index matrix), spatial propagation channels (array-to-user/target spatial propagation), and depolarization channels (polarization-dependent scattering). This unified framework explicitly captures radiation-pattern diversity and polarization reconfigurability, providing a solid basis for clutter-aware ISAC system modeling.

\item \textbf{Joint EM-BB Optimization for Clutter Suppression:} Building on the unified compound channel model, we formulate a mixed-integer nonlinear programming (MINLP) problem to jointly optimize EM- and BB-domain precoders and combiners. Our goal is to maximize radar SCNR while achieving the communication users' SINR requirements and respecting a total transmit-power constraint. To solve this challenging optimization problem, we propose a decomposition-based iterative algorithm utilizing fractional programming (FP), majorization-minimization (MM), second-order cone programming (SOCP), and penalty techniques, ensuring computational tractability and convergence.

\item \textbf{Performance Evaluation and Practical Insights:} Extensive numerical simulations demonstrate significant radar SCNR improvements, up to 11 dB relative to conventional BB-domain beamforming, while consistently meeting communication QoS constraints. Our results highlight the substantial benefits of polarization reconfigurability, demonstrate that low-resolution CRA configurations can closely match high-resolution performance, and offer practical insights on effectively exploiting EM-domain reconfigurability to address clutter challenges in future ISAC deployments.
\end{itemize}

\textit{Notation:} Boldface lower-case and upper-case letters denote column vectors and matrices, respectively. The operators $(\cdot)^{\mathsf{T}}$, $(\cdot)^{\mathsf{H}}$, and $(\cdot)^{-1}$ denote transpose, conjugate transpose, and matrix inverse. For scalar $a$, vector $\mathbf{a}$, and matrix $\mathbf{A}$, $|a|$, $\|\mathbf{a}\|_{1}$, $\|\mathbf{a}\|$, and $\|\mathbf{A}\|_{\mathsf{F}}$ represent the magnitude, $\ell_{1}$ norm, Euclidean norm, and Frobenius norm, respectively. $\mathrm{Tr}\{\mathbf{A}\}$ takes the trace of the
matrix $\mathbf{A}$. The operator $\mathrm{vec}(\mathbf{A})$ stacks the columns of $\mathbf{A}$ into a long column vector. The symbol $\otimes$ denotes the Kronecker product. The notation $\mathrm{blkdiag}\{\mathbf{A}_{1},\mathbf{A}_{2},\ldots,\mathbf{A}_{M}\}$ represents a block-diagonal matrix with blocks $\mathbf{A}_{1}, \mathbf{A}_{2}, \ldots, \mathbf{A}_{M}$, and $\mathrm{diag}\{\mathbf{a}\}$ denotes a diagonal matrix whose diagonal entries are given by the elements of $\mathbf{a}$. The symbols $\mathbf{1}_{M}$ and $\mathbf{I}_{M}$ denote the all-ones vector of length $M$ and the $M \times M$ identity matrix. The operators $\Re\{\cdot\}$ and $\Im\{\cdot\}$ extract the real part and imaginary part of a complex argument, respectively. And $\mathbb{E}\{\cdot\}$ denotes statistical expectation. Finally, $\mathbf{A}(:,j)$ denotes the $j$th column of matrix $\mathbf{A}$.

\section{System Model and Problem Formulation}

\subsection{Single CRA Modeling}

A CRA integrates multiple reconfigurable functionalities within a single hardware platform, enabling dynamic control over two primary EM DoFs: Radiation patterns and polarization states. Radiation-pattern reconfigurability allows CRAs to adaptively steer or shape beams toward intended targets or communication users, effectively mitigating interference. Simultaneously, polarization reconfigurability facilitates switching among distinct polarization modes, significantly improving target-clutter discrimination in sensing applications. The combination of these capabilities provides multidimensional EM-domain adaptation, leading to substantial enhancements in sensing accuracy and communication performance.

\begin{figure*}[!t]
  \centering
  \includegraphics[width= 6.2in]{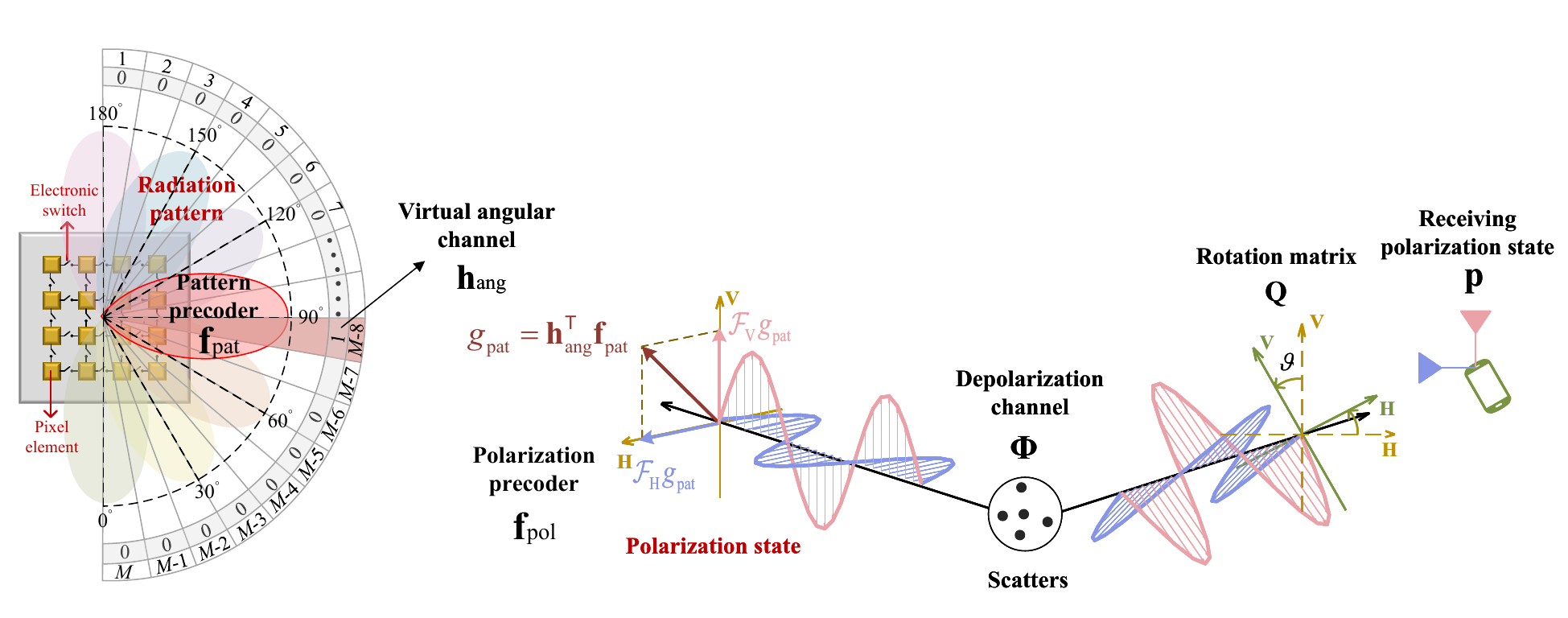}  
  \caption{Illustration of signal propagation in a SISO system with a single CRA.}
  \label{fig:SISO system model}
   \vspace{-0.2 cm}
\end{figure*}

To rigorously quantify the performance gains provided by CRAs, it is essential to develop a comprehensive mathematical model that accurately captures their combined radiation-pattern and polarization characteristics. To facilitate model construction, we initially consider a simplified single-input single-output (SISO)  scenario, wherein a transmitter equipped with a single CRA communicates with a receiver employing a conventional omnidirectional polarized antenna. Fig.~\ref{fig:SISO system model} illustrates the signal propagation for this SISO scenario under the combined radiation-pattern and polarization responses of the CRA. We initially represent the EM-domain responses using separate radiation-pattern and polarization precoders. Subsequently, we derive a unified received-signal model and establish an effective compound channel representation, jointly encapsulating the pattern and polarization effects.

\subsubsection{EM-domain Precoder of Single CRA} To characterize the radiation-pattern response of a CRA, we introduce the \textbf{EM-domain pattern precoder} $\mathbf{f}_{\mathrm{pat}}\in \mathbb{R}^{M}$, representing angular-domain gains across $M$ uniformly sampled spatial directions. Due to practical hardware constraints, the pattern precoder $\mathbf{f}_{\mathrm{pat}}$ is selected from a predefined set $\mathcal{D}_{\mathrm{pat}}$ rather than being freely designed. The gain of the precoder towards a given direction $\theta$ is determined by the \textbf{virtual angular channel} $\mathbf{h}_{\mathrm{ang}}\in\{0,1\}^{M}$, a binary-valued vector containing exactly one nonzero entry whose index corresponds to the angle closest to $\theta$ among the $M$ sampled angles. The effective angular gain in a given direction is thus found by the inner product of the EM-domain pattern precoder $\mathbf{f}_{\mathrm{pat}}$ and the virtual angular channel $\mathbf{h}_{\mathrm{ang}}$:
\begin{equation}\label{eq:g_ang}
g_{\mathrm{pat}} = \mathbf{h}_{\mathrm{ang}}^{\mathsf{T}} \mathbf{f}_{\mathrm{pat}}\in \mathbb{R}.
\end{equation}

Similarly, the polarization response of the CRA is characterized by the \textbf{EM-domain polarization precoder} $\mathbf{f}_{\mathrm{pol}}=[\mathcal{F}_{\mathrm{H}},\mathcal{F}_{\mathrm{V}}]^{\mathsf{T}}\in\mathbb{R}^{2}$, specifying gains for horizontal and vertical polarizations under the normalization constraint $\mathcal{F}_{\mathrm{H}}^{2}+\mathcal{F}_{\mathrm{V}}^{2}=1$. This precoder is also chosen from a predefined polarization codebook $\mathcal{D}_{\mathrm{pol}}$. Accordingly, the resultant polarization gain for the CRA-enabled SISO system is represented as
\begin{equation}\label{eq:g_pol}
g_{\mathrm{pol}} = \mathbf{p}^{\mathsf{T}}\mathbf{Q}\boldsymbol{\Phi}\mathbf{f}_{\mathrm{pol}} \in \mathbb{C},
\end{equation}
where $\boldsymbol{\Phi}\in \mathbb{C}^{2\times 2}$ denotes the depolarization channel matrix, defined by
\begin{equation}\label{eq:Phi}
\boldsymbol{\Phi}=\begin{bmatrix}
\Phi^{\mathrm{HH}} & \Phi^{\mathrm{HV}} \\
\Phi^{\mathrm{VH}} & \Phi^{\mathrm{VV}}
\end{bmatrix},
\end{equation}
with each element representing scattering coefficients between transmitted and received polarizations. Additionally,
\begin{equation}\label{eq:rotation matrix}
\mathbf{Q}=\begin{bmatrix}
\cos\vartheta & \sin\vartheta \\
-\sin\vartheta & \cos\vartheta
\end{bmatrix} \in\mathbb{R}^{2\times 2}
\end{equation}
represents the polarization rotation matrix aligning the local coordinate systems of the transmit and receive antenna polarization states, where $\vartheta$ denotes the antenna co-polarization angle. The polarization state of the receive antenna is represented by $\mathbf{p}=[\mathcal{P}_{\mathrm{H}},\mathcal{P}_{\mathrm{V}}]^{\mathsf{T}}\in\mathbb{R}^{2}$, satisfying $\mathcal{P}_{\mathrm{H}}^{2}+\mathcal{P}_{\mathrm{V}}^{2}=1$.

\subsubsection{Single CRA Channel Model}
Integrating the preceding EM-domain characterizations, the received signal can be formulated as
\begin{equation} \label{eq:received_signal_SISO}
\begin{aligned}
    y &= \alpha \, g_{\mathrm{pol}}  \,  g_{\mathrm{pat}}  \, x + n \\
   &= \alpha \, \mathbf{p}^{\mathsf{T}}\mathbf{Q}\boldsymbol{\Phi} \mathbf{f}_{\mathrm{pol}}\mathbf{h}_{\mathrm{ang}}^{\mathsf{T}} \mathbf{f}_{\mathrm{pat}}  \, x + n,
\end{aligned}
\end{equation}
where $x \in \mathbb{C}$ denotes the transmitted signal, $\alpha$ is the conventional channel coefficient accounting for both attenuation and fading,  and $n\!\sim\!\mathcal{CN}(0,\sigma^{2})$ denotes additive white Gaussian noise (AWGN).

For analytical convenience and to enable a compact optimization formulation, we regroup and simplify the expression in \eqref{eq:received_signal_SISO} as
\begin{equation} \label{eq:received_signal_SISO_transformed}
\begin{aligned}
    y &= \alpha \, \mathbf{p}^{\mathsf{T}}\mathbf{Q}\boldsymbol{\Phi} \mathbf{f}_{\mathrm{pol}}\mathbf{h}_{\mathrm{ang}}^{\mathsf{T}} \mathbf{f}_{\mathrm{pat}}x + n\\
    &= \alpha \,  \mathbf{p}^{\mathsf{T}}\mathbf{Q}\boldsymbol{\Phi} ((\mathbf{h}_{\mathrm{ang}}^{\mathsf{T}} \mathbf{f}_{\mathrm{pat}})\otimes\mathbf{f}_{\mathrm{pol}})x + n\\
    &= \alpha \,  \mathbf{p}^{\mathsf{T}}\mathbf{Q}\boldsymbol{\Phi} (\underbrace{\mathbf{h}_{\mathrm{ang}}\otimes \mathbf{I}_{2}}_{\mathbf{H}_{\mathrm{A}}})^{\mathsf{T}}\underbrace{(\mathbf{f}_{\mathrm{pat}}\otimes\mathbf{f}_{\mathrm{pol}})}_{\mathbf{f}_{\mathrm{EM}}}x + n.\\
\end{aligned}
\end{equation}
The unified EM-domain precoder
\begin{equation}\label{eq:CRA_FEM}
  \mathbf{f}_{\mathrm{EM}}\triangleq \mathbf{f}_{\mathrm{pat}}\otimes \mathbf{f}_{\mathrm{pol}}\in \mathbb{R}^{2M},
\end{equation}
defined in \eqref{eq:received_signal_SISO_transformed} captures joint radiation-pattern and polarization behaviors with the energy constraint  $\|\mathbf{f}_{\mathrm{EM}}\|^{2}=1$, and lies in the set $\mathcal{D}_{\mathrm{EM}}\triangleq\{ \mathbf{f}_{\mathrm{pat}}\otimes \mathbf{f}_{\mathrm{pol}}|\mathbf{f}_{\mathrm{pat}}\in \mathcal{D}_{\mathrm{pat}},\mathbf{f}_{\mathrm{pol}}\in \mathcal{D}_{\mathrm{pol}}\}$. We assume the CRA supports two separable switching networks, so the EM-mode dictionary can be modeled as a Kronecker product. For non-separable switching networks, $\mathcal{D}_{\mathrm{EM}}$ would have to be defined using some general unified dictionary. In \eqref{eq:received_signal_SISO_transformed}, the corresponding extended angular channel is defined by
\begin{equation}\label{eq:CRA_H_A}
\mathbf{H}_{\mathrm{A}}\triangleq \mathbf{h}_{\mathrm{ang}}\otimes \mathbf{I}_{2}\in\{0,1\}^{2M\times 2},
\end{equation}
which replicates the angular indices across two polarization states. The identity matrix $\mathbf{I}_{2}$ is used to retain the CRA’s horizontal and vertical polarization gains in the direction of the intended user, thus facilitating interaction with the subsequent depolarized channel $\boldsymbol{\Phi}$.

Consequently, the received signal in \eqref{eq:received_signal_SISO_transformed} can be rewritten in the following compact form:
\be
\begin{aligned}
y & = \alpha \, \mathbf{p}^{\mathsf{T}} \mathbf{Q}\boldsymbol{\Phi}\mathbf{H}_{\mathrm{A}}^{\mathsf{T}} \mathbf{f}_{\mathrm{EM}} \, x + n \\
& =  \mathbf{p}^{\mathsf{T}} \mathbf{M} \mathbf{f}_{\mathrm{EM}} \, x + n,
\end{aligned}
\ee
where we define the compound channel
\begin{equation} \label{eq:single CRA Heff}
   \mathbf{M} \triangleq \alpha \, \mathbf{Q}\boldsymbol{\Phi} \mathbf{H}_{\mathrm{A}}^{\mathsf{T}}\in \mathbb{C}^{2\times 2M},
\end{equation}
which captures the combined effects of the virtual angular-domain information $\mathbf{H}_{\mathrm{A}}$,  depolarization channel $\boldsymbol{\Phi}$, polarization rotation matrix $\mathbf{Q}$, pathloss and fading $\alpha$.
The above formulation for the EM-domain precoder $\mathbf{f}_{\mathrm{EM}}$ and the compound channel $\mathbf{M}$ for a single CRA lays the foundation for subsequent modeling of CRA arrays.

\begin{figure}[!t]
  \centering
  \includegraphics[width= 3.4in]{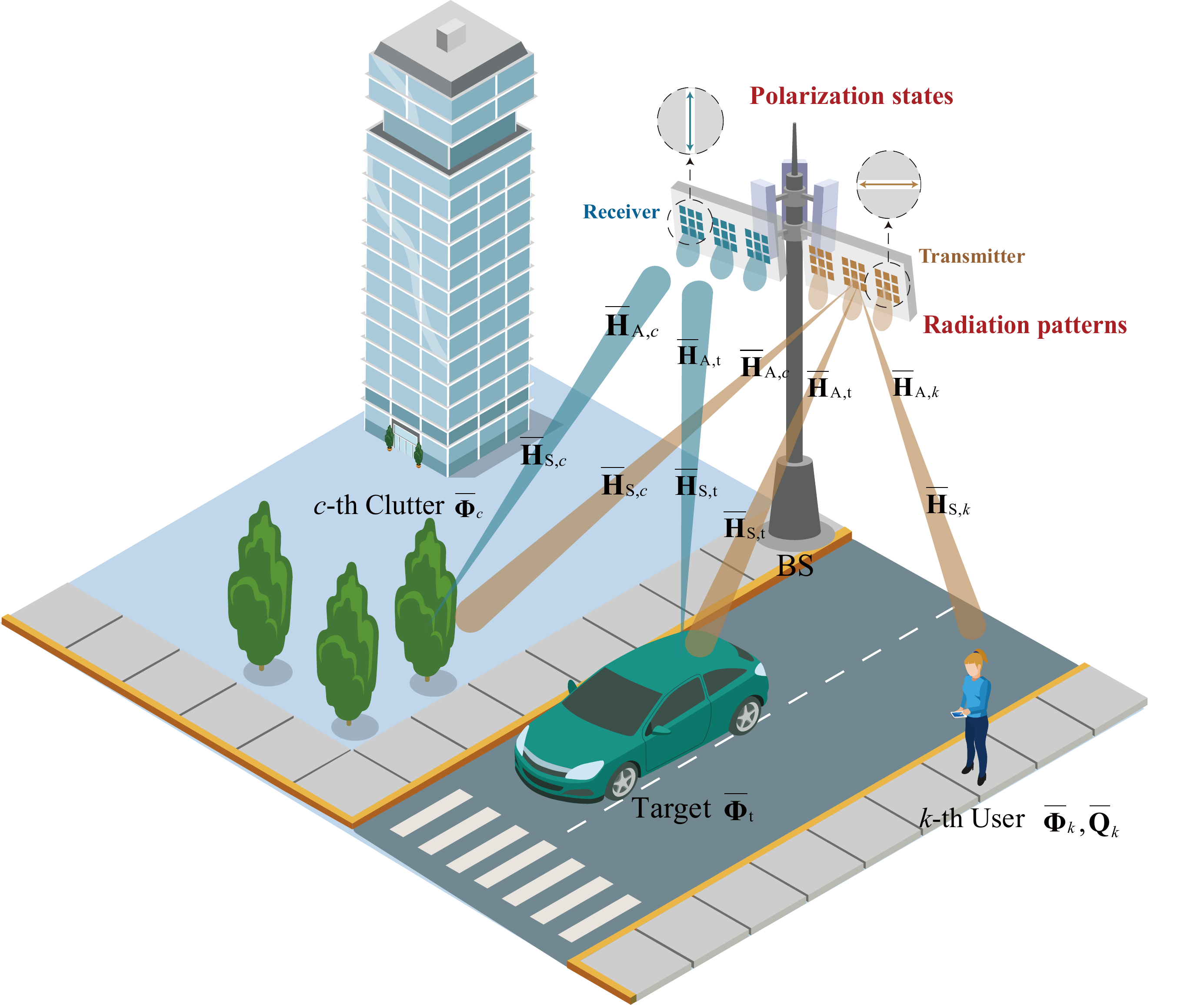}
  \caption{An ISAC system empowered by CRA arrays.}
  \label{fig:systemmodel}
\end{figure}

\subsection{CRA Array Modeling}
The flexibility provided by CRAs can be more fully exploited by integrating them into arrays. CRA arrays offer dynamic reconfigurability of radiation patterns and polarization states in the EM domain, complementing conventional digital beamforming in the BB domain. Such multi-domain adaptability substantially improves the system's ability to distinguish targets from strong clutter and mitigate interference, achieving more accurate and reliable sensing performance in clutter-rich environments.

Fig.~\ref{fig:systemmodel} depicts a CRA-array-empowered ISAC scenario. In this configuration, a base station (BS) simultaneously performs radar sensing of a single target in the presence of $C$ clutter scatterers, while supporting downlink communications to $K$ single-antenna users. The BS is equipped with separate transmit and receive  arrays, each composed of $N$ CRAs with dynamic EM-domain reconfigurability. In contrast, each user employs a conventional omnidirectional polarized antenna. We assume that there are $L$ distinct propagation paths between the BS and each user.

\subsubsection{EM-domain Precoder and Combiner}
To harness the full potential of a CRA array, we establish a detailed mathematical model for both the EM-domain precoder $\mathbf{F}_{\mathrm{EM}}$ at the BS transmitter and the EM-domain combiner $\mathbf{W}_{\mathrm{EM}}$ at the BS receiver. Extending the single-CRA precoder described in \eqref{eq:CRA_FEM}, the array-level EM-domain precoder is structured as
\begin{equation}\label{eq:F_EM_T_extended}
\mathbf{F}_{\mathrm{EM}}\triangleq\mathrm{blkdiag}\{\mathbf{f}_{\mathrm{EM},1}, \mathbf{f}_{\mathrm{EM},2}, \dots, \mathbf{f}_{\mathrm{EM},N}\}\in\mathbb{R}^{2MN\times N},
\end{equation}
where $\mathbf{f}_{\mathrm{EM},n}=\mathbf{f}_{\mathrm{pat},n}\otimes \mathbf{f}_{\mathrm{pol},n}\in \mathbb{R}^{2M}$ represents the EM-domain precoder for the $n$-th transmit antenna. Here, $\mathbf{f}_{\mathrm{pat},n}\in \mathbb{R}^{M}$ captures angular-domain gains across $M$ spatial angles, while $\mathbf{f}_{\mathrm{pol},n}=[\mathcal{F}_{\mathrm{H},n},\mathcal{F}_{\mathrm{V},n}]^{\mathsf{T}}\in \mathbb{R}^{2}$ defines horizontal and vertical polarization gains, satisfying the energy normalization $\mathcal{F}_{\mathrm{H},n}^{2}+\mathcal{F}_{\mathrm{V},n}^2=1$. To ensure uniformity across antennas, we impose the normalization $\|\mathbf{f}_{\mathrm{EM},n}\|^{2}=1,\forall n$.

Similarly, the EM-domain combiner at the receiver is defined as
\begin{equation}\label{eq:F_EM_R_extended}
\mathbf{W}_{\mathrm{EM}}\triangleq\mathrm{blkdiag}\{\mathbf{w}_{\mathrm{EM},1},\mathbf{w}_{\mathrm{EM},2}, \dots,\mathbf{w}_{\mathrm{EM},N}\}\in\mathbb{R}^{2MN\times N},
\end{equation}
where $\mathbf{w}_{\mathrm{EM},n}=\mathbf{w}_{\mathrm{pat},n}\otimes \mathbf{w}_{\mathrm{pol},n}\in \mathbb{R}^{2M}$ represents the EM-domain response of the $n$-th receive antenna. Here, $\mathbf{w}_{\mathrm{pat},n}\in\mathbb{R}^{M}$ denotes angular gains, and $\mathbf{w}_{\mathrm{pol},n}=[\mathcal{W}_{\mathrm{H},n},\mathcal{W}_{\mathrm{V},n}]^{\mathsf{T}}$ specifies the polarization components, similarly normalized as $\mathcal{W}_{\mathrm{H},n}^{2}+\mathcal{W}_{\mathrm{V},n}^{2}=1$ and satisfying $\|\mathbf{w}_{\mathrm{EM},n}\|^{2}=1,\forall n$.

\subsubsection{Compound Channel Model}
Based on the EM-domain precoder $\mathbf{F}_{\mathrm{EM}}$ and combiner $\mathbf{W}_{\mathrm{EM}}$, we construct a compound channel model for the CRA array by extending the single-CRA formulation in \eqref{eq:single CRA Heff}. The proposed model jointly captures three essential components: The virtual angular channel, spatial propagation channel, and depolarization effects with polarization rotation. In the following, we explicitly derive these components and the compound channel by taking the communication user as an example. The corresponding formulations for the radar target and clutter follow analogously.

\begin{table*}[!t]
\caption{Channel Notations}
\begin{center}
\begin{tabular}{ccccc}
   \toprule
    \small Entity    &  \small Virtual Angular  Channel &  \small  Spatial  Channel & \small Depolarization Channel&\small Rotation Matrix \\
    \midrule
     \small User  & \small $\overline{\mathbf{H}}_{\mathrm{A},k,l}\in\{0,1\}^{2MN\times 2N}$ & \small$\overline{\mathbf{H}}_{\mathrm{S},k,l}\in\mathbb{C}^{2N\times 2}$ & \small$\bm{\Phi}_{k,l}\in\mathbb{C}^{2\times 2}$& \small$\mathbf{Q}_{k}\in\mathbb{R}^{2\times 2}$ \\
     \vspace{1mm}
     \small Target   & \small $\overline{\mathbf{H}}_{\mathrm{A},\mathrm{t}}\in\{0,1\}^{2MN\times 2N}$ &  \small$\overline{\mathbf{H}}_{\mathrm{S},\mathrm{t}}\in\mathbb{C}^{2N\times 2}$ & \small$\bm{\Phi}_{\mathrm{t}}\in\mathbb{C}^{2\times 2}$& \textbackslash \\
     \vspace{1mm}
     \small Clutter & \small $\overline{\mathbf{H}}_{\mathrm{A},c}\in\{0,1\}^{2MN\times 2N}$& \small $\overline{\mathbf{H}}_{\mathrm{S},c}\in\mathbb{C}^{2N\times 2}$& \small$\bm{\Phi}_{c}\in\mathbb{C}^{2\times 2}$& \textbackslash  \\
    \bottomrule
\hline
\end{tabular}
\label{table:1}
\end{center}
\vspace{-0.4 cm}
\end{table*}

\begin{table}[!t]
\caption{Compound Channel Models}
\begin{center}
\begin{tabular}{cc}
   \toprule
    \small Entity    &\small Compound Channel \\
    \midrule
     \small User      &\small$\mathbf{M}_{k}\triangleq \sum_{l=1}^{L}\mathbf{Q}_{k}\bm{\Phi}_{k,l}\overline{\mathbf{H}}_{\mathrm{S},k,l}^{\mathsf{H}}\overline{\mathbf{H}}_{\mathrm{A},k,l}^{\mathsf{T}}\in \mathbb{C}^{2\times 2MN}$\\
     \vspace{1mm}
     \small Target   &\small$\mathbf{M}_{\mathrm{t}}\triangleq \mathbf{\overline{H}}_{\mathrm{A},\mathrm{t}}\overline{\mathbf{H}}_{\mathrm{S},\mathrm{t}}
\bm{\Phi}_{\mathrm{t}}
\overline{\mathbf{H}}_{\mathrm{S},\mathrm{t}}^{\mathsf{H}}\mathbf{\overline{H}}_{\mathrm{A},\mathrm{t}}^{\mathsf{T}}\in \mathbb{C}^{2MN\times 2MN}$\\
     \vspace{1mm}
     \small Clutter &\small$\mathbf{M}_{c}\triangleq \mathbf{\overline{H}}_{\mathrm{A},c}\overline{\mathbf{H}}_{\mathrm{S},c}
\bm{\Phi}_{c}
\overline{\mathbf{H}}_{\mathrm{S},c}^{\mathsf{H}}\mathbf{\overline{H}}_{\mathrm{A},c}^{\mathsf{T}}\in \mathbb{C}^{2MN\times 2MN}$\\
    \bottomrule
\hline
\end{tabular}
\label{table:2}
\end{center}
\vspace{-0.4 cm}
\end{table}

\textbf{Virtual Angular Channel:}
Under the far-field assumption, the antenna elements observe each signal arrival from an identical angle. The virtual angular channel vector $\mathbf{h}_{\mathrm{ang},k,l}\in\{0,1\}^{M}$ is a one-hot vector whose nonzero entry specifies the unique angle of the $l$-th propagation path for user $k$. Then, the virtual angular channel vector is duplicated  into the polarization domains, yielding
\begin{equation}
\mathbf{H}_{\mathrm{A},k,l}=\mathbf{h}_{\mathrm{ang},k,l}\otimes \mathbf{I}_{2}\in\{0,1\}^{2M\times 2},~\forall k,l,
\end{equation}
where the inclusion of $\mathbf{I}_{2}$ ensures that the horizontal and vertical polarization gains of each CRA are preserved. Considering all $N$ antenna elements, the complete virtual angular channel for the CRA array becomes:
\begin{equation}\label{eq:H_A_extended}
\overline{\mathbf{H}}_{\mathrm{A},k,l}=\mathbf{I}_{N}\otimes\mathbf{H}_{\mathrm{A},k,l}\in\{0,1\}^{2MN\times 2N},~\forall k,l.
\end{equation}

\textbf{Spatial Channel:}
In addition to fading and path loss, the spatial channel incorporates properties such as inter-element phase shifts. Defining the spatial channel as $\mathbf{h}_{\mathrm{S},k,l}=\alpha_{k,l}\mathbf{a}_{k,l}\in\mathbb{C}^{N}$, where $\alpha_{k,l}$ denotes path loss and $\mathbf{a}_{k,l}$ is the steering vector, we also replicate it across the two polarization dimensions:
\begin{equation}
\label{eq:H_S_extended} \overline{\mathbf{H}}_{\mathrm{S},k,l}= \mathbf{h}_{\mathrm{S},k,l}\otimes \mathbf{I}_{2}\in\mathbb{C}^{2N\times 2},~\forall k,l.
\end{equation}

\textbf{Depolarization Channel:}
The depolarization channel, or polarization scattering matrix, captures polarization alterations due to scattering and multipath propagation. For the $l$-th path of user $k$, we define this matrix as $\boldsymbol{\Phi}_{k,l}$. Additionally, local polarization coordinate misalignment between the BS and users is addressed through a rotation matrix $\mathbf{Q}_{k}\in\mathbb{R}^{2\times 2}$. Note that in radar scenarios where the transmitter and receiver are co-located, this rotation matrix is unnecessary.

With the above definitions of the three core channel components for the communication users, analogous formulations for radar targets and clutter are summarized in Table~\ref{table:1}. Integrating these components, the complete compound channel for user $k$ is succinctly expressed as:
\begin{equation} \label{eq:Heff}
\mathbf{M}_{k}= \sum_{l=1}^{L}\mathbf{Q}_{k}\boldsymbol{\Phi}_{k,l}\overline{\mathbf{H}}_{\mathrm{S},k,l}^{\mathsf{H}}\overline{\mathbf{H}}_{\mathrm{A},k,l}^{\mathsf{T}}\in \mathbb{C}^{2\times 2MN},~\forall k.
\end{equation}
The compound channel models for all considered entities are summarized in Table~\ref{table:2}.

\subsection{ISAC System Model}
\subsubsection{Transmit Signal}
To simultaneously support high-quality communication and precise radar sensing, the transmitted signal is designed as a dual-functional waveform in the BB domain, comprising precoded communication symbols and dedicated radar probing signals:
\setcounter{equation}{16}
\begin{equation}\label{eq:dual-functional waveform}
\mathbf{x}_{\mathrm{BB}} = \mathbf{F}_{\mathrm{BB},\mathrm{c}}\mathbf{s}_{\mathrm{c}} + \mathbf{F}_{\mathrm{BB},\mathrm{r}}\mathbf{s}_{\mathrm{r}} = \mathbf{F}_{\mathrm{BB}}\mathbf{s},
\end{equation}
where $\mathbf{s}_{\mathrm{c}}\in \mathbb{C}^{K}$ is the communication symbol vector with covariance $\mathbb{E}\{\mathbf{s}_{\mathrm{c}}\mathbf{s}_{\mathrm{c}}^{\mathsf{H}}\}=\mathbf{I}_{K}$, and $\mathbf{s}_{\mathrm{r}}\in \mathbb{C}^{N-K}$ denotes the radar probing signal satisfying $\mathbb{E}\{\mathbf{s}_{\mathrm{r}}\mathbf{s}_{\mathrm{r}}^{\mathsf{H}}\}=\mathbf{I}_{N-K}$. These two signal vectors are statistically independent. The matrices $\mathbf{F}_{\mathrm{BB},\mathrm{c}}\in \mathbb{C}^{N\times K}$ and $\mathbf{F}_{\mathrm{BB},\mathrm{r}}\in \mathbb{C}^{N\times (N-K)}$ are digital precoders dedicated to communication and radar sensing, respectively. We further define the integrated symbol vector $\mathbf{s}\triangleq[\mathbf{s}_{\mathrm{c}}^{\mathsf{T}}, \mathbf{s}_{\mathrm{r}}^{\mathsf{T}}]^{\mathsf{T}}\in \mathbb{C}^{N}$ and the composite digital precoder as $\mathbf{F}_{\mathrm{BB}}\triangleq[\mathbf{F}_{\mathrm{BB},\mathrm{c}},\mathbf{F}_{\mathrm{BB},\mathrm{r}}]\triangleq[\mathbf{f}_{\mathrm{BB},1}, \mathbf{f}_{\mathrm{BB},2}, \ldots, \mathbf{f}_{\mathrm{BB},N}]\in \mathbb{C}^{N\times N}$, where $\mathbf{f}_{\mathrm{BB},j}$ is the $j$-th column of $\mathbf{F}_{\mathrm{BB}}$.

Following digital precoding in the BB domain, the resultant signal is further shaped by the EM-domain precoder $\mathbf{F}_{\mathrm{EM}}$, yielding the final transmitted signal as
\begin{equation}
\mathbf{x}_{\mathrm{EM}} = \mathbf{F}_{\mathrm{EM}}\mathbf{x}_{\mathrm{BB}},
\end{equation}
which is subsequently radiated by the CRA array.

\subsubsection{Received Signal at the Communication Users}
Based on the compound channel model derived in \eqref{eq:Heff}, the received signal at the $k$-th user can be expressed as
\begin{equation}\label{eq:received signal User}
y_{k} =\mathbf{p}_{k}^{\mathsf{T}}\mathbf{M}_{k}\mathbf{F}_{\mathrm{EM}}\mathbf{F}_{\mathrm{BB}}\mathbf{s}+n_{k},~\forall k,
\end{equation}
where $\mathbf{p}_{k}\triangleq [\mathcal{P}_{\mathrm{H},k},\mathcal{P}_{\mathrm{V},k}]^{\mathsf{T}}\in\mathbb{R}^{2}$ represents the horizontal and vertical polarization gain components at user $k$ satisfying $\mathcal{P}_{\mathrm{H},k}^{2}+\mathcal{P}_{\mathrm{V},k}^{2}=1$, and $n_{k}\!\sim\!\mathcal{CN}(0,\sigma^{2}_{k})$ denotes AWGN at the $k$-th user.
Consequently, the SINR at user $k$ is calculated as
\begin{equation}\label{eq:SINR User simplified}
\begin{aligned}
\mathrm{SINR}_{k}&=\frac{|\mathbf{p}_{k}^{\mathsf{T}}\mathbf{M}_{k}\mathbf{F}_{\mathrm{EM}}\mathbf{f}_{\mathrm{BB},k}|^{2}}{\sum_{j\neq k}^{N}|\mathbf{p}_{k}^{\mathsf{T}}
\mathbf{M}_{k}\mathbf{F}_{\mathrm{EM}}\mathbf{f}_{\mathrm{BB},j}|^{2}+\sigma_{k}^{2}}, ~\forall k.\\
\end{aligned}
\end{equation}

\subsubsection{Received Radar Sensing Signal}
For radar sensing, the dual-functional waveform transmitted by the CRA array illuminates both the target and the surrounding clutter.
The EM-domain precoder $\mathbf{F}_{\mathrm{EM}}$ determines the transmit radiation pattern and polarization state, while the virtual angular channels $\overline{\mathbf{H}}_{\mathrm{A},\mathrm{t}}^{\mathsf{T}}$ and $\overline{\mathbf{H}}_{\mathrm{A},c}^{\mathsf{T}}$ characterize the directions to the target and the $c$-th clutter patch, respectively.
The signals propagate through forward spatial channels $\overline{\mathbf{H}}_{\mathrm{S},\mathrm{t}}^{\mathsf{H}}$ and $\overline{\mathbf{H}}_{\mathrm{S},c}^{\mathsf{H}}$, undergo polarization transformation described by depolarization matrices $\boldsymbol{\Phi}_{\mathrm{t}}$ and $\boldsymbol{\Phi}_{c}$, and return via backward spatial channels $\overline{\mathbf{H}}_{\mathrm{S},\mathrm{t}}$ and $\overline{\mathbf{H}}_{\mathrm{S},c}$ and virtual angular channels $\overline{\mathbf{H}}_{\mathrm{A},\mathrm{t}}$ and $\overline{\mathbf{H}}_{\mathrm{A},c}$. The received echoes are then captured and processed by the EM-domain combiner $\mathbf{W}_{\mathrm{EM}}$ at the BS.

Combining these effects, the received echo at the BS can be compactly expressed as:
\begin{equation}\label{eq:received signal sensing}
\begin{aligned}
\mathbf{y}_{\mathrm{r}}
&=
\mathbf{W}_{\mathrm{EM}}^{\mathsf{T}}
\overline{\mathbf{H}}_{\mathrm{A},\mathrm{t}}
\overline{\mathbf{H}}_{\mathrm{S},\mathrm{t}}
\boldsymbol{\Phi}_{\mathrm{t}}
\overline{\mathbf{H}}_{\mathrm{S},\mathrm{t}}^{\mathsf{H}}
\overline{\mathbf{H}}_{\mathrm{A},\mathrm{t}}^{\mathsf{T}}
\mathbf{F}_{\mathrm{EM}}
\mathbf{F}_{\mathrm{BB}}\mathbf{s}  \\
&
+\sum_{c=1}^{C}
\mathbf{W}_{\mathrm{EM}}^{\mathsf{T}}
\overline{\mathbf{H}}_{\mathrm{A},c}
\overline{\mathbf{H}}_{\mathrm{S},c}
\boldsymbol{\Phi}_{c}
\overline{\mathbf{H}}_{\mathrm{S},c}^{\mathsf{H}}
\overline{\mathbf{H}}_{\mathrm{A},c}^{\mathsf{T}}
\mathbf{F}_{\mathrm{EM}}
\mathbf{F}_{\mathrm{BB}}\mathbf{s}
+\mathbf{n}_{\mathrm{r}},
\end{aligned}
\end{equation}
where $\mathbf{n}_{\mathrm{r}}\!\sim\!\mathcal{CN}(0,\sigma_{\mathrm{r}}^{2}\mathbf{I}_{N})$ is AWGN at the BS. The received echo can be succinctly described using the compound channels $\mathbf{M}_{\mathrm{t}}$ and $\mathbf{M}_{c}$ defined in Table~II:
\begin{equation}
\label{eq:received signal sensing concise}
\mathbf{y}_{\mathrm{r}}
=
\mathbf{W}_{\mathrm{EM}}^{\mathsf{T}}
\mathbf{M}_{\mathrm{t}}
\mathbf{F}_{\mathrm{EM}}\mathbf{F}_{\mathrm{BB}}\mathbf{s}
+
\sum_{c=1}^{C}
\mathbf{W}_{\mathrm{EM}}^{\mathsf{T}}
\mathbf{M}_{c}
\mathbf{F}_{\mathrm{EM}}\mathbf{F}_{\mathrm{BB}}\mathbf{s}
+
\mathbf{n}_{\mathrm{r}}.
\end{equation}
To further enhance the target echo and suppress clutter, the received vector $\mathbf{y}_{\mathrm{r}}$ is processed by a digital combiner $\mathbf{w}_{\mathrm{BB}}\in\mathbb{C}^{N}$. The processed radar signal output is then given by
\begin{equation}\label{eq:received signal sensing filter}
\begin{aligned}
y_{\mathrm{r}}
= \mathbf{w}_{\mathrm{BB}}^{\mathsf{H}} \mathbf{y}_{\mathrm{r}} =&\
\mathbf{w}_{\mathrm{BB}}^{\mathsf{H}}
\mathbf{W}_{\mathrm{EM}}^{\mathsf{T}}
\mathbf{M}_{\mathrm{t}}
\mathbf{F}_{\mathrm{EM}}
\mathbf{F}_{\mathrm{BB}}\mathbf{s} \\
&+
\mathbf{w}_{\mathrm{BB}}^{\mathsf{H}}
\sum_{c=1}^{C}
\mathbf{W}_{\mathrm{EM}}^{\mathsf{T}}
\mathbf{M}_{c}
\mathbf{F}_{\mathrm{EM}}
\mathbf{F}_{\mathrm{BB}}\mathbf{s}
+
\mathbf{w}_{\mathrm{BB}}^{\mathsf{H}}\mathbf{n}_{\mathrm{r}}.
\end{aligned}
\end{equation}
The associated radar output SCNR follows directly from \eqref{eq:received signal sensing filter} and is given in \eqref{eq:SCNR}  shown at the bottom of this page.

\newcounter{TempEqCnt}
\setcounter{TempEqCnt}{\value{equation}}
\setcounter{equation}{23}
\begin{figure*}[b]
\hrulefill
\begin{equation}\label{eq:SCNR}
\begin{aligned}
\mathrm{SCNR}_{\mathrm{r}}=&\frac{\mathbb{E}\{\mathbf{w}_{\mathrm{BB}}^{\mathsf{H}}\mathbf{W}_{\mathrm{EM}}^{\mathsf{T}}\mathbf{M}_{\mathrm{t}}\mathbf{F}_{\mathrm{EM}}\mathbf{F}_{\mathrm{BB}}\mathbf{s}\mathbf{s}^{\mathsf{H}}
\mathbf{F}_{\mathrm{BB}}^{\mathsf{H}}\mathbf{F}_{\mathrm{EM}}^{\mathsf{T}}\mathbf{M}_{\mathrm{t}}^{\mathsf{H}}\mathbf{W}_{\mathrm{EM}}\mathbf{w}_{\mathrm{BB}}\}}
{\mathbb{E}\{\mathbf{w}_{\mathrm{BB}}^{\mathsf{H}}(\sum_{c=1}^{C}\mathbf{W}_{\mathrm{EM}}^{\mathsf{T}}\mathbf{M}_{c}\mathbf{F}_{\mathrm{EM}}\mathbf{F}_{\mathrm{BB}}\mathbf{s}\mathbf{s}^{\mathsf{H}}
\mathbf{F}_{\mathrm{BB}}^{\mathsf{H}}\mathbf{F}_{\mathrm{EM}}^{\mathsf{T}}\mathbf{M}_{c}^{\mathsf{H}}\mathbf{W}_{\mathrm{EM}})\mathbf{w}_{\mathrm{BB}}\}+\mathbb{E}\{\mathbf{w}_{\mathrm{BB}}^{\mathsf{H}}\mathbf{n}_{\mathrm{r}}\mathbf{n}_{\mathrm{r}}^{\mathsf{H}}\mathbf{w}_{\mathrm{BB}}\}}.
\end{aligned}
\end{equation}
\begin{equation}\label{eq:SCNR_transformed}
\begin{aligned}
\mathrm{SCNR}_{\mathrm{r}}=&\frac{\mathrm{Tr}\{(\mathbf{F}_{\mathrm{BB}}\otimes\mathbf{w}_{\mathrm{BB}}^{\ast})^{\mathsf{T}}(\mathbf{F}_{\mathrm{EM}}\otimes\mathbf{W}_{\mathrm{EM}})^{\mathsf{T}}
\mathbb{E}\{\mathbf{m}_{\mathrm{t}}\mathbf{m}_{\mathrm{t}}^{\mathsf{H}}\}(\mathbf{F}_{\mathrm{EM}}\otimes\mathbf{W}_{\mathrm{EM}})^{\ast}(\mathbf{F}_{\mathrm{BB}}\otimes\mathbf{w}_{\mathrm{BB}}^{\ast})^{\ast}\}}
{\mathrm{Tr}\{(\mathbf{F}_{\mathrm{BB}}\otimes\mathbf{w}_{\mathrm{BB}}^{\ast})^{\mathsf{T}}(\mathbf{F}_{\mathrm{EM}}\otimes\mathbf{W}_{\mathrm{EM}})^{\mathsf{T}}
\mathbb{E}\{\sum_{c=1}^{C}\mathbf{m}_{c}\mathbf{m}_{c}^{\mathsf{H}}\}(\mathbf{F}_{\mathrm{EM}}\otimes\mathbf{W}_{\mathrm{EM}})^{\ast}(\mathbf{F}_{\mathrm{BB}}\otimes\mathbf{w}_{\mathrm{BB}}^{\ast})^{\ast}\}
+\sigma_{\mathrm{r}}^{2}\mathbf{w}_{\mathrm{BB}}^{\mathsf{H}}\mathbf{w}_{\mathrm{BB}}}.
\end{aligned}
\end{equation}
\end{figure*}

In practice, the instantaneous channel for the target and clutter are generally unavailable, necessitating the use of statistical channel information. To effectively design precoders and combiners using statistical characteristics, the radar SCNR expression must be explicitly reformulated based on these statistics. Therefore, we derive the equivalent statistical SCNR formulation shown in \eqref{eq:SCNR_transformed} at the bottom of this page. The equivalent compound channel vectors $\mathbf{m}_{\mathrm{t}}$  and $\mathbf{m}_{c}$ corresponding to the target and clutter, respectively, are introduced for convenience as follows:
\setcounter{equation}{25}
\begin{equation}\label{eq:compound_channel_vectors}
\begin{aligned}
\mathbf{m}_{\mathrm{t}} &\triangleq \mathrm{vec}(\mathbf{M}_{\mathrm{t}})=\widetilde{\mathbf{H}}_{\mathrm{A},\mathrm{t}}\widetilde{\mathbf{H}}_{\mathrm{S},\mathrm{t}}\pmb{\phi}_{\mathrm{t}}\in \mathbb{C}^{4M^{2}N^{2}}, \\
\mathbf{m}_{c} &\triangleq \mathrm{vec}(\mathbf{M}_{c})=\widetilde{\mathbf{H}}_{\mathrm{A},c}\widetilde{\mathbf{H}}_{\mathrm{S},c}\pmb{\phi}_{c}\in \mathbb{C}^{4M^{2}N^{2}},  ~\forall c,\\
\widetilde{\mathbf{H}}_{\mathrm{A},\mathrm{t}} &\triangleq \overline{\mathbf{H}}_{\mathrm{A},\mathrm{t}}\otimes\overline{\mathbf{H}}_{\mathrm{A},\mathrm{t}}\in \mathbb{C}^{4M^{2}N^{2}\times 4N^{2}}, \\
\widetilde{\mathbf{H}}_{\mathrm{A},c} &\triangleq \overline{\mathbf{H}}_{\mathrm{A},c}\otimes\overline{\mathbf{H}}_{\mathrm{A},c}\in \mathbb{C}^{4M^{2}N^{2}\times 4N^{2}}, ~\forall c,\\
\widetilde{\mathbf{H}}_{\mathrm{S},\mathrm{t}} &\triangleq \overline{\mathbf{H}}_{\mathrm{S},\mathrm{t}}^{\ast}\otimes\overline{\mathbf{H}}_{\mathrm{S},\mathrm{t}}\in \mathbb{C}^{4N^{2}\times 4}, \\
\widetilde{\mathbf{H}}_{\mathrm{S},c} &\triangleq \overline{\mathbf{H}}_{\mathrm{S},c}^{\ast}\otimes\overline{\mathbf{H}}_{\mathrm{S},c}\in \mathbb{C}^{4N^{2}\times 4},  ~\forall c,\\
\pmb{\phi}_{\mathrm{t}} &\triangleq \mathrm{vec}(\boldsymbol{\Phi}_{\mathrm{t}})\in \mathbb{C}^{4}, \\
\pmb{\phi}_{c} &\triangleq \mathrm{vec}(\boldsymbol{\Phi}_{c})\in \mathbb{C}^{4}, ~\forall c.
\end{aligned}
\end{equation}
Therefore, the radar SCNR can be expressed compactly in terms of the channel statistics as follows:
\begin{equation}\label{eq:SCNR_transformed_final}
\mathrm{SCNR}_{\mathrm{r}}=\frac{\mathrm{Tr}\{\mathbf{B}^{\mathsf{T}} \mathbf{E}^{\mathsf{T}}\boldsymbol{\Omega}_{\mathrm{t}}\mathbf{E}^{\ast}\mathbf{B}^{\ast}\}}
{\mathrm{Tr}\{\mathbf{B}^{\mathsf{T}} \mathbf{E}^{\mathsf{T}}\boldsymbol{\Omega}_{\mathrm{c}}\mathbf{E}^{\ast}\mathbf{B}^{\ast}\}+\sigma_{\mathrm{r}}^{2}\mathbf{w}_{\mathrm{BB}}^{\mathsf{H}}\mathbf{w}_{\mathrm{BB}}},
\end{equation}
where $\mathbf{B}$ and $\mathbf{E}$ correspond to the BB- and EM-domain transceiver designs, respectively, defined as:
\begin{equation}
\begin{aligned}
\mathbf{B} &\triangleq \mathbf{F}_{\mathrm{BB}}\otimes\mathbf{w}_{\mathrm{BB}}^{\ast}\in \mathbb{C}^{N^{2}\times N}, \\
\mathbf{E} &\triangleq \mathbf{F}_{\mathrm{EM}}\otimes\mathbf{W}_{\mathrm{EM}}\in \mathbb{R}^{4M^{2}N^{2}\times N^{2}}.
\end{aligned}
\end{equation}
The covariance matrices of the compound channels for the target and clutter  $\boldsymbol{\Omega}_{\mathrm{t}}$ and $\boldsymbol{\Omega}_{\mathrm{c}}$, are defined as:
\begin{equation}
\begin{aligned}
\label{covariance compound channel}
\boldsymbol{\Omega}_{\mathrm{t}}&\triangleq \mathbb{E}\{\mathbf{m}_{\mathrm{t}}\mathbf{m}_{\mathrm{t}}^{\mathsf{H}}\}
=\mathbf{\widetilde{H}}_{\mathrm{A},\mathrm{t}}\mathbf{\widetilde{H}}_{\mathrm{S},\mathrm{t}}\boldsymbol{\Sigma}_{\mathrm{t}}\mathbf{\widetilde{H}}_{\mathrm{S},\mathrm{t}}^{\mathsf{H}}\mathbf{\widetilde{H}}_{\mathrm{A},\mathrm{t}}^{\mathsf{T}}, \\
\boldsymbol{\Omega}_{\mathrm{c}}&\triangleq \mathbb{E}\Big\{\sum_{c=1}^{C}\mathbf{m}_{c}\mathbf{m}_{c}^{\mathsf{H}}\Big\}
=\sum_{c=1}^{C}\mathbf{\widetilde{H}}_{\mathrm{A},c}\mathbf{\widetilde{H}}_{\mathrm{S},c}\boldsymbol{\Sigma}_{c}\mathbf{\widetilde{H}}_{\mathrm{S},c}^{\mathsf{H}}\mathbf{\widetilde{H}}_{\mathrm{A},c}^{\mathsf{T}},
\end{aligned}
\end{equation}
where $\boldsymbol{\Sigma}_{\mathrm{t}}\triangleq\mathbb{E}\{\pmb{\phi}_{\mathrm{t}}\pmb{\phi}_{\mathrm{t}}^{\mathsf{H}}\}$ and
$\boldsymbol{\Sigma}_{c}\triangleq\mathbb{E}\{\pmb{\phi}_{c}\pmb{\phi}_{c}^{\mathsf{H}}\}$ represent the covariance matrices of the target and clutter depolarization channels, respectively.

In many radar scenarios, the angles and ranges of the targets and clutter either remain static or change slowly over time. Thus, the transformed virtual angular channels $\mathbf{\widetilde{H}}_{\mathrm{A},\mathrm{t}}$, $\mathbf{\widetilde{H}}_{\mathrm{A},c}$ and spatial channels $\mathbf{\widetilde{H}}_{\mathrm{S},\mathrm{t}}$, $\mathbf{\widetilde{H}}_{\mathrm{S},c}$ can be considered quasi-static within a single sensing frame. Conversely, the depolarization channel vectors $\pmb{\phi}_{\mathrm{t}}$, $\pmb{\phi}_{c}$ experience significant fluctuations due to their sensitivity to object orientation, surface characteristics, and small-scale scattering. Hence, modeling these depolarization vectors as random variables is reasonable.

The covariance matrices $\boldsymbol{\Omega}_{\mathrm{t}}$ and $\boldsymbol{\Omega}_{\mathrm{c}}$ of the compound channels defined in \eqref{covariance compound channel} are both estimated directly from the received data using sample averaging under predefined CRA radiation and polarization configurations. In particular, the clutter covariance $\boldsymbol{\Omega}_{\mathrm{c}}$ is estimated from clutter-only snapshots, exploiting the stationary nature of the background. For the target, continuous tracking enables reliable estimation of its covariance matrix $\boldsymbol{\Omega}_{\mathrm{t}}$ based on accumulated historical target-present observations  \cite{R.  Liu clutter}.


\subsection{Problem Formulation}
Building on the compound channel model derived above, we now formulate an optimization framework to fully exploit the joint EM- and BB-domain design capabilities of CRA arrays. Our objective is to jointly optimize the EM-domain precoder $\mathbf{F}_{\mathrm{EM}}$ and combiner $\mathbf{W}_{\mathrm{EM}}$, and the BB-domain precoder $\mathbf{F}_{\mathrm{BB}}$ and combiner $\mathbf{w}_{\mathrm{BB}}$ to maximize the radar SCNR, subject to communication SINR constraints $\epsilon_{k}$, discrete EM-domain configurations constrained by the finite set $\mathcal{D}_{\mathrm{EM}}$, and a total transmit power constraint $P_{\mathrm{T}}$. This design goal can be formulated as the following mixed-integer nonlinear programming (MINLP) problem:
\begin{subequations}
\label{eq:original_problem}
\begin{align}
\label{eq:original_problem_a}
\max_{\mathbf{F}_{\mathrm{EM}},\mathbf{W}_{\mathrm{EM}}, \mathbf{F}_{\mathrm{BB}},\mathbf{w}_{\mathrm{BB}}}&
~\mathrm{SCNR}_{\mathrm{r}}\\
\mathrm{s.t.}\ \ \ \ \ \quad
\label{eq:original_problem_b}
&\mathrm{SINR}_{k}\geq \epsilon_{k},~ \forall k,\\
\label{eq:original_problem_c}
&\mathbf{f}_{\mathrm{EM},n},\mathbf{w}_{\mathrm{EM},n}\in \mathcal{D}_{\mathrm{EM}},~\forall n,\\
\label{eq:original_problem_d}
&\|\mathbf{F}_{\mathrm{BB}}\|_{\mathsf{F}}^{2}\leq P_{\mathrm{T}},
\end{align}
\end{subequations}
where $\mathcal{D}_{\mathrm{EM}}$ represents the finite set of candidate EM-domain modes available for each individual CRA element, as detailed in Section II-A.

This MINLP problem is inherently challenging due to the fractional SCNR objective \eqref{eq:original_problem_a}, fractional SINR constraints \eqref{eq:original_problem_b}, and discrete selection constraints \eqref{eq:original_problem_c}. To efficiently handle these challenges, we next propose a decomposition-based iterative algorithm, which partitions the original MINLP into several tractable subproblems and solves them iteratively.

\section{Joint EM- and BB-Domain Designs}
In this section, we first reformulate the original MINLP optimization problem into an equivalent mixed-binary nonlinear program. Subsequently, we propose an alternating optimization algorithm in which each resulting subproblem is iteratively solved using FP, MM, SOCP, and penalty-based techniques. The detailed algorithmic steps are presented as follows.

\subsection{Reformulation of the Original Problem}
A major challenge in designing the EM-domain precoder $\mathbf{F}_{\mathrm{EM}}$ and combiner $\mathbf{W}_{\mathrm{EM}}$ arises due to the discrete mode selection from a predefined finite set $\mathcal{D}_{\mathrm{EM}}$, as indicated by constraint \eqref{eq:original_problem_c}. Directly addressing these discrete variables is computationally demanding. To simplify the optimization process, we introduce an equivalent formulation using dictionary matrices and binary selection variables, thus converting the EM-domain precoder and combiner design into a more tractable mixed-binary nonlinear programming form.

Specifically, the EM-domain precoder $\mathbf{F}_{\mathrm{EM}}$ and combiner $\mathbf{W}_{\mathrm{EM}}$ are reformulated as
\begin{align}
\label{eq:FEM_decomposition_T}
\mathbf{F}_{\mathrm{EM}}&=\mathrm{diag}\{\mathrm{vec}(\mathbf{D}_{\mathrm{EM}}\mathbf{S}_{\mathrm{F}})\}\mathbf{C},\\
\label{eq:FEM_decomposition_R}
\mathbf{W}_{\mathrm{EM}}&=\mathrm{diag}\{\mathrm{vec}(\mathbf{D}_{\mathrm{EM}}\mathbf{S}_{\mathrm{W}})\}\mathbf{C},
\end{align}
where $\mathbf{D}_{\mathrm{EM}}$ represents an EM mode dictionary constructed as $\mathbf{D}_{\mathrm{EM}}=\mathbf{D}_{\mathrm{pat}}\otimes\mathbf{D}_{\mathrm{pol}}$. The sub-dictionaries $\mathbf{D}_{\mathrm{pat}}\in\mathbb{R}^{M\times P_{\mathrm{pat}}}$ and $\mathbf{D}_{\mathrm{pol}}\in\mathbb{R}^{2\times P_{\mathrm{pol}}}$ are derived from the radiation pattern set $\mathcal{D}_{\mathrm{pat}}$ and polarization set $\mathcal{D}_{\mathrm{pol}}$, respectively. The sub-dictionary $\mathbf{D}_{\mathrm{pat}}$ contains $P_{\mathrm{pat}}$ candidate radiation patterns across $M$ angular directions, whereas $\mathbf{D}_{\mathrm{pol}}$ includes $P_{\mathrm{pol}}$ candidate polarization states spanning horizontal and vertical orientations. Let $P=P_{\mathrm{pat}}P_{\mathrm{pol}}$ denote the total number of EM modes, so the dimension of the EM mode dictionary is $\mathbf{D}_{\mathrm{EM}}\in\mathbb{R}^{2M\times P}$. We further introduce binary selection matrices $\mathbf{S}_{\mathrm{F}}\in\{0,1\}^{P\times N}$ and $\mathbf{S}_{\mathrm{W}}\in\{0,1\}^{P\times N}$ to explicitly represent the selection of EM modes for the transmit and receive antennas, respectively. Each column $\mathbf{S}_{\mathrm{F}}(:,n)$ and $\mathbf{S}_{\mathrm{W}}(:,n)$ has exactly one nonzero entry corresponding to the selected mode for the $n$-th antenna, ensuring $\|\mathbf{S}_{\mathrm{F}}(:,n)\|_{1}=1$ and $\|\mathbf{S}_{\mathrm{W}}(:,n)\|_{1}=1$. The constant matrix $\mathbf{C}\triangleq\mathbf{I}_{N}\otimes\mathbf{1}_{2M}$ enforces the block-diagonal structure of $\mathbf{F}_{\mathrm{EM}}$ and $\mathbf{W}_{\mathrm{EM}}$, as previously established in \eqref{eq:F_EM_T_extended} and \eqref{eq:F_EM_R_extended}.

With these definitions, the original optimization problem \eqref{eq:original_problem} is transformed into the following equivalent mixed-binary nonlinear programming form:
\begin{subequations}
\label{eq:original_problem_S}
\begin{align}
\label{eq:original_problem_S_a}
\max_{\mathbf{S}_{\mathrm{F}},\mathbf{S}_{\mathrm{W}},\mathbf{F}_{\mathrm{BB}}, \mathbf{w}_{\mathrm{BB}}}&\mathrm{SCNR}_{\mathrm{r}}\\
\mathrm{s.t.}\ \ \ \ \ \
\label{eq:original_problem_S_b}
&\mathrm{SINR}_{k}\geq\epsilon_{k},~\forall k, \\
\label{eq:original_problem_S_c}
&\mathbf{S}_{\mathrm{F}}\in\{0,1\}^{P\times N}, \\
\label{eq:original_problem_S_d}
&\|\mathbf{S}_{\mathrm{F}}(:,n)\|_{1}=1,~\forall n, \\
\label{eq:original_problem_S_e}
&\mathbf{S}_{\mathrm{W}}\in\{0,1\}^{P\times N}, \\
\label{eq:original_problem_S_f}
&\|\mathbf{S}_{\mathrm{W}}(:,n)\|_{1}=1,~\forall n, \\
\label{eq:original_problem_S_g}
&\|\mathbf{F}_{\mathrm{BB}}\|_{\mathsf{F}}^{2}\leq P_{\mathrm{T}}.
\end{align}
\end{subequations}
In the following, we decompose it into four subproblems, each dedicated to the optimization of a specific variable, thereby enabling a more tractable solution framework.

\subsection{EM-domain Precoder Design}
For a fixed digital precoder $\mathbf{F}_{\mathrm{BB}}$ and combiner $\mathbf{w}_{\mathrm{BB}}$, and given the receiver-side EM mode selection matrix $\mathbf{S}_{\mathrm{W}}$, the subproblem for designing the transmitter-side EM mode selection matrix $\mathbf{S}_{\mathrm{F}}$ can be written as
\begin{subequations}
\label{eq:original_problem_SEM_T}
\begin{align}
\label{eq:original_problem_SEM_T_a}
    \max_{\mathbf{S}_{\mathrm{F}}}\ \ \ &\mathrm{SCNR}_{\mathrm{r}}\\
     \mathrm{s.t.}\ \ \
     \label{eq:original_problem_SEM_T_b}
    &\mathrm{SINR}_{k}\geq \epsilon_{k},~\forall k, \\
    \label{eq:original_problem_SEM_T_c}
    &\mathbf{S}_{\mathrm{F}}\in\{0,1\}^{P\times N}, \\
    \label{eq:original_problem_SEM_T_d}
    &\|\mathbf{S}_{\mathrm{F}}(:,n)\|_{1}=1,~\forall n.
\end{align}
\end{subequations}

\subsubsection{Objective Function Transformation}
We rewrite the objective function in \eqref{eq:original_problem_SEM_T_a} as
\begin{equation}\label{eq:SCNR_FEM_T}
\begin{aligned}
\mathrm{SCNR}_{\mathrm{r}}=&\frac{\overline{\mathbf{f}}_{\mathrm{EM}}^{\mathsf{T}}\overline{\mathbf{E}}_{\mathrm{F}}\overline{\mathbf{f}}_{\mathrm{EM}}}
{\overline{\mathbf{f}}_{\mathrm{EM}}^{\mathsf{T}}\overline{\mathbf{E}}_{\mathrm{F},\mathrm{c}}\overline{\mathbf{f}}_{\mathrm{EM}}
+\sigma_{\mathrm{r}}^{2}\mathbf{w}_{\mathrm{BB}}^{\mathsf{H}}\mathbf{w}_{\mathrm{BB}}},
\end{aligned}
\end{equation}
with
\begin{equation}
\begin{aligned}
  \overline{\mathbf{f}}_{\mathrm{EM}} &\triangleq \mathrm{vec}(\mathbf{F}_{\mathrm{EM}}^{\mathsf{T}}),
  \\
   \overline{\mathbf{E}}_{\mathrm{F}} &\triangleq \mathbf{E}_{\mathrm{F}}^{\mathsf{T}}(\boldsymbol{\Omega}_{\mathrm{t}}\otimes\mathbf{B}\mathbf{B}^{\mathsf{H}})\mathbf{E}_{\mathrm{F}}, \\
   \overline{\mathbf{E}}_{\mathrm{F},\mathrm{c}} &\triangleq \mathbf{E}_{\mathrm{F}}^{\mathsf{T}}(\boldsymbol{\Omega}_{\mathrm{c}}\otimes\mathbf{B}\mathbf{B}^{\mathsf{H}})\mathbf{E}_{\mathrm{F}}, \\
  \mathbf{E}_{\mathrm{F}} &\triangleq (\mathbf{I}_{2MN}\otimes\mathbf{K}_{2MN,N}\otimes\mathbf{I}_{N})(\mathbf{I}_{2MN^{2}}\otimes\mathrm{vec}(\mathbf{W}_{\mathrm{EM}}^{\mathsf{T}})).\\
\end{aligned}
\end{equation}
The commutation matrix $\mathbf{K}_{P,Q}$ is a permutation matrix defined as $\mathbf{K}_{P,Q}=\sum_{i=1}^{P}\sum_{j=1}^{Q}(\mathbf{e}_{i}\mathbf{e}_{j}^{\mathsf{T}})\otimes(\mathbf{e}_{j}\mathbf{e}_{i}^{\mathsf{T}})$, where $\mathbf{e}_{i}$ and $\mathbf{e}_{j}$ denote the canonical basis vectors.
Then, due to the non-convexity of the single-ratio fractional objective function in \eqref{eq:SCNR_FEM_T}, the FP technique is employed to convert it into an equivalent form, which yields
\begin{equation}\label{eq:SCNR_FEM_T FP}
\begin{aligned}
\overline{\mathbf{f}}_{\mathrm{EM}}^{\mathsf{T}}\overline{\mathbf{E}}_{\mathrm{F}}\overline{\mathbf{f}}_{\mathrm{EM}}
-\gamma(\overline{\mathbf{f}}_{\mathrm{EM}}^{\mathsf{T}}\overline{\mathbf{E}}_{\mathrm{F},\mathrm{c}}\overline{\mathbf{f}}_{\mathrm{EM}}
+\sigma_{\mathrm{r}}^{2}\mathbf{w}_{\mathrm{BB}}^{\mathsf{H}}\mathbf{w}_{\mathrm{BB}}),
\end{aligned}
\end{equation}
where the auxiliary variable $\gamma$ has the optimal value $\gamma^{\star}=\mathrm{SCNR}_{\mathrm{r}}$ \cite{FP}.
However, it can be observed that the first quadratic term $\overline{\mathbf{f}}_{\mathrm{EM}}^{\mathsf{T}}\overline{\mathbf{E}}_{\mathrm{F}}\overline{\mathbf{f}}_{\mathrm{EM}}$ in \eqref{eq:SCNR_FEM_T FP} is non-concave. To address this, we utilize the MM method to relax it into a linear surrogate function \cite{MM}. Consequently, the final transformed concave objective function for optimizing  the transmitter-side EM mode selection matrix $\mathbf{S}_{\mathrm{F}}$ can be written as follows
\begin{equation}\label{eq:SCNR_FEM_T_final}
\begin{aligned}
f_{\mathrm{SCNR}}(\mathbf{S}_{\mathrm{F}})&\triangleq
(\overline{\mathbf{f}}_{\mathrm{EM}}^{(t)})^{\mathsf{T}}\overline{\mathbf{E}}_{\mathrm{F}}\overline{\mathbf{f}}_{\mathrm{EM}}^{(t)}
\!+\!2\Re\{(\overline{\mathbf{f}}_{\mathrm{EM}}^{(t)})^{\mathsf{T}}\overline{\mathbf{E}}_{\mathrm{F}}(\overline{\mathbf{f}}_{\mathrm{EM}}\!-\!\overline{\mathbf{f}}_{\mathrm{EM}}^{(t)})\}
\\
&\ \ \ -\gamma(\overline{\mathbf{f}}_{\mathrm{EM}}^{\mathsf{T}}\overline{\mathbf{E}}_{\mathrm{F},\mathrm{c}}\overline{\mathbf{f}}_{\mathrm{EM}}
+\sigma_{\mathrm{r}}^{2}\mathbf{w}_{\mathrm{BB}}^{\mathsf{H}}\mathbf{w}_{\mathrm{BB}}).
\end{aligned}
\end{equation}

\subsubsection{SINR Constraint Transformations}
It is evident that the communication SINR constraint in \eqref{eq:original_problem_SEM_T_b} is also a fractional non-convex function. To handle this difficulty, we adopt a conservative SOCP-based approximation following \cite{SOCP}, through which the original SINR constraint is replaced by a tractable second-order cone constraint, yielding
\begin{equation} \label{eq:SINRUE_FEM_T_final}
\begin{aligned}
&f_{\mathrm{u},k}(\mathbf{S}_{\mathrm{F}})
\triangleq \Re\{\mathbf{p}_{k}^{\mathsf{T}}\mathbf{M}_{k}\mathbf{F}_{\mathrm{EM}}\mathbf{f}_{\mathrm{BB},k}\}\\
&\ \ \ \ -\sqrt{\epsilon_{k}}\Big(\sum_{j\neq k}^{N}|\mathbf{p}_{k}^{\mathsf{T}}
\mathbf{M}_{k}\mathbf{F}_{\mathrm{EM}}\mathbf{f}_{\mathrm{BB},j}|^{2}+\sigma_{k}^{2}\Big)^{\frac{1}{2}}\geq 0,~\forall k.\\
\end{aligned}
\end{equation}

\subsubsection{Boolean Constraint Transformation}
We now address the Boolean constraint in \eqref{eq:original_problem_SEM_T_c}, following the strategy in \cite{Boolean}. We first reformulate it as a quadratic maximization problem with a box constraint,
\begin{subequations}
\label{eq:Boolean_SEM_T}
\begin{align}
    \label{eq:Boolean_SEM_T_a}
    \max_{\mathbf{S}_{\mathrm{F}}}&\ \ \mathbf{s}_{\mathrm{F}}^{\mathsf{T}}(\mathbf{s}_{\mathrm{F}}-\mathbf{1}_{PN})\\
    \mathrm{s.t. }\label{eq:Boolean_SEM_T_b}
     &\ \ \mathbf{S}_{\mathrm{F}}\in[0,1]^{P\times N},
\end{align}
\end{subequations}
where $\mathbf{s}_{\mathrm{F}}\triangleq\mathrm{vec}(\mathbf{S}_{\mathrm{F}})\in\mathbb{R}^{PN}$ is introduced for brevity and the binary variable $\mathbf{S}_{\mathrm{F}}$ is relaxed to be continuous.
It can be observed that the objective function in \eqref{eq:Boolean_SEM_T_a} is a non-concave quadratic term, and thus we employ the MM approach to convert it into a linear form at the current point $\mathbf{s}_{\mathrm{F}}^{(t)}$, as follows:
\begin{equation}
\begin{aligned}
  \label{eq:Boolean_SEM_T_MM}
  \mathbf{s}_{\mathrm{F}}^{\mathsf{T}}(\mathbf{s}_{\mathrm{F}}-\mathbf{1}_{PN})\geq2(\mathbf{s}_{\mathrm{F}}^{(t)})^{\mathsf{T}}\mathbf{s}_{\mathrm{F}}-(\mathbf{s}_{\mathrm{F}}^{(t)})^{\mathsf{T}}\mathbf{s}_{\mathrm{F}}^{(t)} - \mathbf{s}_{\mathrm{F}}^{\mathsf{T}}\mathbf{1}_{PN}.
\end{aligned}
\end{equation}
Then, we incorporate \eqref{eq:Boolean_SEM_T_MM} as a penalty term in the objective function \eqref{eq:SCNR_FEM_T_final} with a box constraint \eqref{eq:Boolean_SEM_T_b}, leading to
\begin{subequations}
\label{eq:Boolean_SEM_T_objectivefunction}
\begin{align}
    \label{eq:Boolean_SEM_T_objectivefunction_a}
    \max_{\mathbf{S}_{\mathrm{F}}}&\ \  f_{\mathrm{SCNR}}(\mathbf{S}_{\mathrm{F}})+ \varrho_{1}(2\mathbf{s}_{\mathrm{F}}^{(t)}-\mathbf{1}_{PN})^{\mathsf{T}}\mathbf{s}_{\mathrm{F}}\\
    \mathrm{s.t. }\label{eq:Boolean_SEM_T_objectivefunction_b}
     &\ \ \mathbf{S}_{\mathrm{F}}\in[0,1]^{P\times N},
\end{align}
\end{subequations}
where $\varrho_{1}$ is the penalty parameter introduced to promote binary selection behavior.

Finally, by incorporating the QoS constraint in \eqref{eq:SINRUE_FEM_T_final} and the linear constraint on EM mode selection in \eqref{eq:original_problem_SEM_T_d}, the subproblem for optimizing the transmitter-side EM mode selection matrix $\mathbf{S}_{\mathrm{F}}$ in the EM domain is formulated as follows
\begin{subequations}
\label{eq:original_problem_SEM_T_final}
\begin{align}
\label{eq:original_problem_SEM_T_final_a}
    \max_{\mathbf{S}_{\mathrm{F}}}&\ \ f_{\mathrm{SCNR}}(\mathbf{S}_{\mathrm{F}})+ \varrho_{1}(2\mathbf{s}_{\mathrm{F}}^{(t)}-\mathbf{1}_{PN})^{\mathsf{T}}\mathbf{s}_{\mathrm{F}}\\
     \mathrm{s.t.}\
     \label{eq:original_problem_SEM_T_final_b}
    &\ \ f_{\mathrm{u},k}(\mathbf{S}_{\mathrm{F}})\geq 0,~\forall k, \\
    \label{eq:original_problem_SEM_T_final_c}
    &\ \ \mathbf{S}_{\mathrm{F}}\in[0,1]^{P\times N},\\
    \label{eq:original_problem_SEM_T_final_d}
    &\ \ \|\mathbf{S}_{\mathrm{F}}(:,n)\|_{1}=1,~\forall n,
\end{align}
\end{subequations}
which can be efficiently tackled by standard convex optimization tools. Once the optimal selection matrix $\mathbf{S}_{\mathrm{F}}$ is obtained, the corresponding  EM-domain precoder   $\mathbf{F}_{\mathrm{EM}}$ can be constructed according to \eqref{eq:FEM_decomposition_T}.

\subsection{ EM-domain Combiner Design}
Given the other variables $\mathbf{F}_{\mathrm{BB}}$, $\mathbf{w}_{\mathrm{BB}}$, and $\mathbf{S}_{\mathrm{F}}$, we obtain the following subproblem for determining the optimal receiver-side EM mode selection matrix $\mathbf{S}_{\mathrm{W}}$:
\begin{subequations}
\label{eq:original_problem_SEM_R}
\begin{align}
\label{eq:original_problem_SEM_R_a}
    \max_{\mathbf{S}_{\mathrm{W}}}\ \ \ &\mathrm{SCNR}_{\mathrm{r}}\\
     \mathrm{s.t.}\ \ \
    \label{eq:original_problem_SEM_R_b}
    &\mathbf{S}_{\mathrm{W}}\in\{0,1\}^{P\times N}, \\
    \label{eq:original_problem_SEM_R_c}
    &\|\mathbf{S}_{\mathrm{W}}(:,n)\|_{1}=1,~\forall n.
\end{align}
\end{subequations}

\subsubsection{Objective Function Transformation}
Similar to the approach for the EM-domain precoder selection matrix in Section III.B, we represent the objective in \eqref{eq:original_problem_SEM_R_a} as a function of the variable $\mathbf{S}_{\mathrm{W}}$, which yields
\begin{equation}\label{eq:SCNR_FEM_R}
\begin{aligned}
\mathrm{SCNR}_{\mathrm{r}}=&\frac{\overline{\mathbf{w}}_{\mathrm{EM}}^{\mathsf{T}}\overline{\mathbf{E}}_{\mathrm{W}}\overline{\mathbf{w}}_{\mathrm{EM}}}
{\overline{\mathbf{w}}_{\mathrm{EM}}^{\mathsf{T}}\overline{\mathbf{E}}_{\mathrm{W},\mathrm{c}}\overline{\mathbf{w}}_{\mathrm{EM}}
+\sigma_{\mathrm{r}}^{2}\mathbf{w}_{\mathrm{BB}}^{\mathsf{H}}\mathbf{w}_{\mathrm{BB}}},
\end{aligned}
\end{equation}
with
\begin{equation}
\begin{aligned}
  \overline{\mathbf{w}}_{\mathrm{EM}} &\triangleq \mathrm{vec}(\mathbf{W}_{\mathrm{EM}}^{\mathsf{T}}),\\
  \overline{\mathbf{E}}_{\mathrm{W}} &\triangleq \mathbf{E}_{\mathrm{W}}^{\mathsf{T}}(\boldsymbol{\Omega}_{\mathrm{t}}\otimes\mathbf{B}\mathbf{B}^{\mathsf{H}})\mathbf{E}_{\mathrm{W}}, \\
  \overline{\mathbf{E}}_{\mathrm{W},\mathrm{c}} &\triangleq \mathbf{E}_{\mathrm{W}}^{\mathsf{T}}(\boldsymbol{\Omega}_{\mathrm{c}}\otimes\mathbf{B}\mathbf{B}^{\mathsf{H}})\mathbf{E}_{\mathrm{W}}, \\
\mathbf{E}_{\mathrm{W}} &\triangleq (\mathbf{I}_{2MN}\otimes\mathbf{K}_{2MN,N}\otimes\mathbf{I}_{N})(\mathrm{vec}(\mathbf{F}_{\mathrm{EM}}^{\mathsf{T}})\otimes\mathbf{I}_{2MN^{2}}).
\end{aligned}
\end{equation}
Following the algorithm development in Section III-B, the objective can be equivalently reformulated using the following concave surrogate
\begin{equation}\label{eq:SCNR_FEM_R_final}
\begin{aligned}
&f_{\mathrm{SCNR}}(\mathbf{S}_{\mathrm{W}})\triangleq\\
&(\overline{\mathbf{w}}_{\mathrm{EM}}^{(t)})^{\mathsf{T}}\overline{\mathbf{E}}_{\mathrm{W}}\overline{\mathbf{w}}_{\mathrm{EM}}^{(t)}
+2\Re\{(\overline{\mathbf{w}}_{\mathrm{EM}}^{(t)})^{\mathsf{T}}\overline{\mathbf{E}}_{\mathrm{W}}(\overline{\mathbf{w}}_{\mathrm{EM}}-\overline{\mathbf{w}}_{\mathrm{EM}}^{(t)})\}
\\
&-\gamma\Big(\overline{\mathbf{w}}_{\mathrm{EM}}^{\mathsf{T}}\overline{\mathbf{E}}_{\mathrm{W},\mathrm{c}}\overline{\mathbf{w}}_{\mathrm{EM}}
+\sigma_{\mathrm{r}}^{2}\mathbf{w}_{\mathrm{BB}}^{\mathsf{H}}\mathbf{w}_{\mathrm{BB}}\Big),
\end{aligned}
\end{equation}
where $\gamma$ is the auxiliary variable whose optimal value satisfies $\gamma^{\star}=\mathrm{SCNR}_{\mathrm{r}}$.

\subsubsection{Boolean Constraint Transformation}
Again, as with the penalty-MM-based algorithm in Section III-B, we handle the Boolean constraint by relaxing the binary variable $\mathbf{S}_{\mathrm{W}}$ to be continuous and incorporating a penalty term into the objective function. As a result, the original Boolean constraint in \eqref{eq:original_problem_SEM_R_b} is replaced by the following optimization problem:
\begin{subequations}
\label{eq:Boolean_SEM_R_objectivefunction}
\begin{align}
    \label{eq:Boolean_SEM_R_objectivefunction_a}
    \max_{\mathbf{S}_{\mathrm{W}}}&\ \  f_{\mathrm{SCNR}}(\mathbf{S}_{\mathrm{W}})+ \varrho_{2}(2\mathbf{s}_{\mathrm{W}}^{(t)}-\mathbf{1}_{PN})^{\mathsf{T}}\mathbf{s}_{\mathrm{W}}\\
    \mathrm{s.t. }\label{eq:Boolean_SEM_R_objectivefunction_b}
     &\ \ \mathbf{S}_{\mathrm{W}}\in[0,1]^{P\times N},
\end{align}
\end{subequations}
where $\mathbf{s}_{\mathrm{W}}\triangleq\mathrm{vec}(\mathbf{S}_{\mathrm{W}})\in\mathbb{R}^{PN}$ and $\varrho_{2}$ is the penalty parameter.

Finally, incorporating the concave objective function in \eqref{eq:SCNR_FEM_R_final} and the transformed Boolean constraint in \eqref{eq:Boolean_SEM_R_objectivefunction} along with the linear mode selection constraint outlined in \eqref{eq:original_problem_SEM_R_c}, the subproblem for optimizing the receiver-side EM mode selection matrix $\mathbf{S}_{\mathrm{W}}$ is expressed as
\begin{subequations}
\label{eq:original_problem_SEM_R_final}
\begin{align}
\label{eq:original_problem_SEM_R_final_a}
    \max_{\mathbf{S}_{\mathrm{W}}}&\ \ f_{\mathrm{SCNR}}(\mathbf{S}_{\mathrm{W}})+ \varrho_{2}(2\mathbf{s}_{\mathrm{W}}^{(t)}-\mathbf{1}_{PN})^{\mathsf{T}}\mathbf{s}_{\mathrm{W}}\\
     \mathrm{s.t.}\
     \label{eq:original_problem_SEM_R_final_b}
    &\ \ \mathbf{S}_{\mathrm{W}}\in[0,1]^{P\times N}, \\
    \label{eq:original_problem_SEM_R_final_c}
    &\ \ \|\mathbf{S}_{\mathrm{W}}(:,n)\|_{1}=1,~\forall n,
\end{align}
\end{subequations}
which can be solved using standard convex optimization tools. The optimal selection matrix $\mathbf{S}_{\mathrm{W}}$ allows the construction of the  EM-domain combiner   $\mathbf{W}_{\mathrm{EM}}$ as per \eqref{eq:FEM_decomposition_R}.

\subsection{Digital Precoder Design}
After designing the  EM-domain precoder $\mathbf{F}_{\mathrm{EM}}$ and combiner $\mathbf{W}_{\mathrm{EM}}$, we shift our focus to the design of the digital precoder in the BB domain. The corresponding subproblem is formulated as follows
\begin{subequations}
\label{eq:original_problem_W}
\begin{align}
\label{eq:original_problem_W_a}
    \max_{\mathbf{F}_{\mathrm{BB}}}&\ \ \mathrm{SCNR}_{\mathrm{r}}\\
     \mathrm{s.t.}
     \label{eq:original_problem_W_b}
    &\ \ \mathrm{SINR}_{k}\geq \epsilon_{k}, \forall k, \\
    \label{eq:original_problem_W_c}
    &\ \ \|\mathbf{F}_{\mathrm{BB}}\|_{\mathsf{F}}^{2}\leq P_{\mathrm{T}}.
\end{align}
\end{subequations}

\subsubsection{Objective Function Transformation}
To facilitate the subsequent transformation of \eqref{eq:original_problem_W_a} with respect to $\mathbf{F}_{\mathrm{BB}}$, we rewrite it in the equivalent form
\begin{equation}\label{eq:SCNR_W}
\begin{aligned}
\mathrm{SCNR}_{\mathrm{r}}=&\frac{\overline{\mathbf{f}}_{\mathrm{BB}}^{\mathsf{H}}\overline{\mathbf{B}}_{\mathrm{F}}\overline{\mathbf{f}}_{\mathrm{BB}}}
{\overline{\mathbf{f}}_{\mathrm{BB}}^{\mathsf{H}}\overline{\mathbf{B}}_{\mathrm{F},\mathrm{c}}\overline{\mathbf{f}}_{\mathrm{BB}}
+\sigma_{\mathrm{r}}^{2}\mathbf{w}_{\mathrm{BB}}^{\mathsf{H}}\mathbf{w}_{\mathrm{BB}}},
\end{aligned}
\end{equation}
with the constant matrices defined as
\begin{equation}
\begin{aligned}
  \overline{\mathbf{f}}_{\mathrm{BB}} &\triangleq \mathrm{vec}(\mathbf{F}_{\mathrm{BB}}^{\mathsf{T}}), 
  \\
  \overline{\mathbf{B}}_{\mathrm{F}}&\triangleq\mathbf{B}_{\mathrm{F}}^{\mathsf{H}}(\mathbf{E}^{\mathsf{T}}\boldsymbol{\Omega}_{\mathrm{t}}^{\mathsf{T}}\mathbf{E}\otimes\mathbf{I}_{N})\mathbf{B}_{\mathrm{F}},\\
 \overline{ \mathbf{B}}_{\mathrm{F},\mathrm{c}}&\triangleq\mathbf{B}_{\mathrm{F}}^{\mathsf{H}}(\mathbf{E}^{\mathsf{T}}\boldsymbol{\Omega}_{\mathrm{c}}^{\mathsf{T}}\mathbf{E}\otimes\mathbf{I}_{N})\mathbf{B}_{\mathrm{F}},\\
  \mathbf{B}_{\mathrm{F}} &\triangleq (\mathbf{I}_{N}\otimes\mathbf{K}_{N,N})(\mathbf{I}_{N^{2}}\otimes\mathbf{w}_{\mathrm{BB}}^{\ast}).
\end{aligned}
\end{equation}
Then, we can leverage the FP-MM-based technique previously developed in Sections III-B and III-C to tackle the fractional non-concave SCNR expression in \eqref{eq:SCNR_W}, which yields
\begin{equation}\label{eq:SCNR_W_final}
\begin{aligned}
f_{\mathrm{SCNR}}(\mathbf{F}_{\mathrm{BB}})\triangleq&\ (\overline{\mathbf{f}}_{\mathrm{BB}}^{(t)})^{\mathsf{H}}\overline{\mathbf{B}}_{\mathrm{F}}\overline{\mathbf{f}}_{\mathrm{BB}}^{(t)}
\!+\!2\Re\{(\overline{\mathbf{f}}_{\mathrm{BB}}^{(t)})^{\mathsf{H}}\overline{\mathbf{B}}_{\mathrm{F}}(\overline{\mathbf{f}}_{\mathrm{BB}}\!-\!\overline{\mathbf{f}}_{\mathrm{BB}}^{(t)})\}\\
&-\gamma(\overline{\mathbf{f}}_{\mathrm{BB}}^{\mathsf{H}}\overline{\mathbf{B}}_{\mathrm{F},\mathrm{c}}\overline{\mathbf{f}}_{\mathrm{BB}}
+\sigma_{\mathrm{r}}^{2}\mathbf{w}_{\mathrm{BB}}^{\mathsf{H}}\mathbf{w}_{\mathrm{BB}}). \\
\end{aligned}
\end{equation}

\subsubsection{Communication SINR Constraint Transformation}
Similar to the approach adopted in \eqref{eq:SINRUE_FEM_T_final}, the non-convex SINR constraint for communication in \eqref{eq:original_problem_W_b} can also be transformed into SOCP form as
\begin{equation} \label{eq:SINRUE_W_final}
f_{\mathrm{u},k}(\mathbf{F}_{\mathrm{BB}})\geq 0,~\forall k,
\end{equation}
with the imaginary-part constraint $\Im\{\overline{p}_{k}^{\mathsf{T}}\mathbf{M}_{k}\mathbf{F}_{\mathrm{EM}}\mathbf{f}_{\mathrm{BB},k}\}=0$.

With both the objective function \eqref{eq:SCNR_W_final} and the SINR constraints \eqref{eq:SINRUE_W_final} now cast in convex forms, the optimization problem with respect to digital precoder $\mathbf{F}_{\mathrm{BB}}$ is written as
\begin{subequations}
\label{eq:original_problem_W_final}
\begin{align}
\label{eq:original_problem_W_final_a}
    \max_{\mathbf{F}_{\mathrm{BB}}}&\ \ f_{\mathrm{SCNR}}(\mathbf{F}_{\mathrm{BB}})\\
     \mathrm{s.t.}
     \label{eq:original_problem_W_final_b}
    &\ \ f_{\mathrm{u},k}(\mathbf{F}_{\mathrm{BB}})\geq 0, \forall k, \\
    &\ \ \Im\{\overline{p}_{k}^{\mathsf{T}}\mathbf{M}_{k}\mathbf{F}_{\mathrm{EM}}\mathbf{f}_{\mathrm{BB},k}\}=0,~\forall k,\\
    \label{eq:original_problem_W_final_c}
    &\ \ \|\mathbf{F}_{\mathrm{BB}}\|_{\mathsf{F}}^{2}\leq P_{\mathrm{T}}.
\end{align}
\end{subequations}
As this convex optimization problem can be efficiently solved by various standard algorithms, the detailed solution procedure is omitted here.

\subsection{Digital Combiner Design}
For the design of the digital combiner $\mathbf{w}_{\mathrm{BB}}$, note that only the objective function \eqref{eq:original_problem_S_a} in the original problem \eqref{eq:original_problem_S} involves $\mathbf{w}_{\mathrm{BB}}$. Therefore, the resulting optimization subproblem can be simply stated as
\begin{equation}
\label{eq:original_problem_u}
    \max_{\mathbf{w}_{\mathrm{BB}}}\ \ \frac{\mathbf{w}_{\mathrm{BB}}^{\mathsf{H}}\overline{\mathbf{B}}_{\mathrm{W},\mathrm{t}}\mathbf{w}_{\mathrm{BB}}}{\mathbf{w}_{\mathrm{BB}}^{\mathsf{H}}\overline{\mathbf{B}}_{\mathrm{W},\mathrm{c}}\mathbf{w}_{\mathrm{BB}}},
\end{equation}
where we define
\begin{equation}
\begin{aligned}
  \overline{\mathbf{B}}_{\mathrm{W},\mathrm{t}} &\triangleq \mathbf{B}_{\mathrm{W}}^{\mathsf{H}}(\mathbf{E}^{\mathsf{T}}\boldsymbol{\Omega}_{\mathrm{t}}\mathbf{E}\otimes\mathbf{I}_{N})\mathbf{B}_{\mathrm{W}},\\
   \overline{\mathbf{B}}_{\mathrm{W},\mathrm{c}} &\triangleq \mathbf{B}_{\mathrm{W}}^{\mathsf{H}}(\mathbf{E}^{\mathsf{T}}\boldsymbol{\Omega}_{\mathrm{c}}\mathbf{E}\otimes\mathbf{I}_{N})\mathbf{B}_{\mathrm{W}}+\sigma_{\mathrm{r}}^{2}\mathbf{I}_{N},\\
   \mathbf{B}_{\mathrm{W}} &\triangleq (\mathbf{I}_{N}\otimes\mathbf{K}_{N,N})(\mathrm{vec}(\mathbf{F}_{\mathrm{BB}}^{\mathsf{H}})\otimes\mathbf{I}_{N}).
\end{aligned}
\end{equation}
The matrices $\overline{\mathbf{B}}_{\mathrm{W},\mathrm{t}}$ and $\overline{\mathbf{B}}_{\mathrm{W},\mathrm{c}}$ are Hermitian, and $\mathbf{B}_{\mathrm{W},\mathrm{c}}$ is positive definite. Hence, problem \eqref{eq:original_problem_u} is a generalized Rayleigh quotient optimization whose optimal solution is known to be the eigenvector of $\mathbf{B}_{\mathrm{W},\mathrm{c}}^{-1}\mathbf{B}_{\mathrm{W},\mathrm{t}}$ corresponding to the largest eigenvalue.

\subsection{Summary and Complexity Analysis}
\subsubsection{Summary}
Based on the preceding derivations, the proposed joint EM- and BB-domain design algorithm is summarized in Algorithm 1. With proper initialization, the transmitter-side EM mode selection matrix $\mathbf{S}_{\mathrm{F}}$, the receiver-side EM mode selection matrix $\mathbf{S}_{\mathrm{W}}$, the digital precoder $\mathbf{F}_{\mathrm{BB}}$, the digital combiner $\mathbf{w}_{\mathrm{BB}}$, and the auxiliary variable $\gamma$ are iteratively updated until the algorithm converges.


\begin{algorithm}[t]
\caption{The proposed joint EM- and BB-domain design for CRA-array-empowered  ISAC systems.}
\label{alg:SH}
    \begin{algorithmic}[1]
    \begin{small}
    \REQUIRE $\mathbf{p}_{k},\mathbf{Q}_{k},\bm{\Phi}_{k,l},
\overline{\mathbf{H}}_{\mathrm{S},k,l},\mathbf{\overline{H}}_{\mathrm{A},k,l}, \forall k, l$, \\
\ \ \ \ \ $\boldsymbol{\Omega}_{\mathrm{t}}$, $\boldsymbol{\Omega}_{\mathrm{c}}$, $\mathbf{D}_{\mathrm{EM}}$.\\
    \ENSURE $\mathbf{F}_{\mathrm{EM}}$, $\mathbf{W}_{\mathrm{EM}}$, $\mathbf{F}_{\mathrm{BB}}$, $\mathbf{w}_{\mathrm{BB}}$.
            \STATE {Initialize $\mathbf{F}_{\mathrm{EM}}$, $\mathbf{W}_{\mathrm{EM}}$, $\mathbf{F}_{\mathrm{BB}}$, $\mathbf{w}_{\mathrm{BB}}$.}
            \WHILE {no convergence of $\mathbf{F}_{\mathrm{EM}}$, $\mathbf{W}_{\mathrm{EM}}$, $\mathbf{F}_{\mathrm{BB}}$, $\mathbf{w}_{\mathrm{BB}}$}
                   \STATE {Update the transmitter-side EM mode selection matrix $\mathbf{S}_{\mathrm{F}}$ by solving \eqref{eq:original_problem_SEM_T_final}.}
                   \STATE {Construct the EM-domain precoder   $\mathbf{F}_{\mathrm{EM}}$ by \eqref{eq:FEM_decomposition_T}.}
                   \STATE {Update the receiver-side EM mode selection matrix $\mathbf{S}_{\mathrm{W}}$ by solving \eqref{eq:original_problem_SEM_R_final}.}
                   \STATE {Construct the EM-domain combiner   $\mathbf{W}_{\mathrm{EM}}$ by \eqref{eq:FEM_decomposition_R}.}
                   \STATE {Update the BB-domain digital precoder $\mathbf{F}_{\mathrm{BB}}$ by solving \eqref{eq:original_problem_W_final}.}
                   \STATE {Update the BB-domain digital combiner $\mathbf{w}_{\mathrm{BB}}$ by solving \eqref{eq:original_problem_u}.}
                   \STATE {Update the auxiliary variable $\gamma^{\star}=\mathrm{SCNR}_{\mathrm{r}}$.}
            \ENDWHILE
            \STATE {Return $\mathbf{F}_{\mathrm{EM}}$, $\mathbf{W}_{\mathrm{EM}}$, $\mathbf{F}_{\mathrm{BB}}$, $\mathbf{w}_{\mathrm{BB}}$.}
    \end{small}
    \vspace{-0.0 cm}
    \end{algorithmic}
    \vspace{-0.0 cm}
\end{algorithm}

\subsubsection{Complexity Analysis}
The primary computational cost associated with Algorithm~1 arises from solving the quadratically constrained quadratic program (QCQP) sub-problems \eqref{eq:original_problem_SEM_T_final}, \eqref{eq:original_problem_SEM_R_final} and \eqref{eq:original_problem_W_final} for updating  $\mathbf{S}_{\mathrm{F}}$, $\mathbf{S}_{\mathrm{W}}$, and $\mathbf{F}_{\mathrm{BB}}$, respectively, and from performing the eigenvalue decomposition in \eqref{eq:original_problem_u} to update $\mathbf{w}_{\mathrm{BB}}$. Assuming a generic interior-point method is employed to solve the optimization problems, the computational complexity of solving problem \eqref{eq:original_problem_SEM_T_final} with an $PN$-dimensional variable, problem \eqref{eq:original_problem_SEM_R_final} with an $PN$-dimensional variable, and problem \eqref{eq:original_problem_W_final} with an $N^{2}$-dimensional variable is of order $\mathcal{O}\{P^{3.5}N^{3.5}\}$, $\mathcal{O}\{P^{3.5}N^{3.5}\}$ , $\mathcal{O}\{N^{7}\}$ respectively. Solving for the variable $\mathbf{w}_{\mathrm{BB}}$ in problem \eqref{eq:original_problem_u} requires an eigenvalue decomposition, whose computational complexity is $\mathcal{O}\{N^{3}\}$.
Thus, the computational complexity for one iteration of Algorithm~1 is of order $\mathcal{O}\{2P^{3.5}N^{3.5}+N^{7}+N^{3}\}$ \cite{Lectures 2001}.

\section{Simulation Results}
\begin{figure}[!t]
  \centering
  \vspace{-0.2 cm}
  \includegraphics[width= 3.5 in]{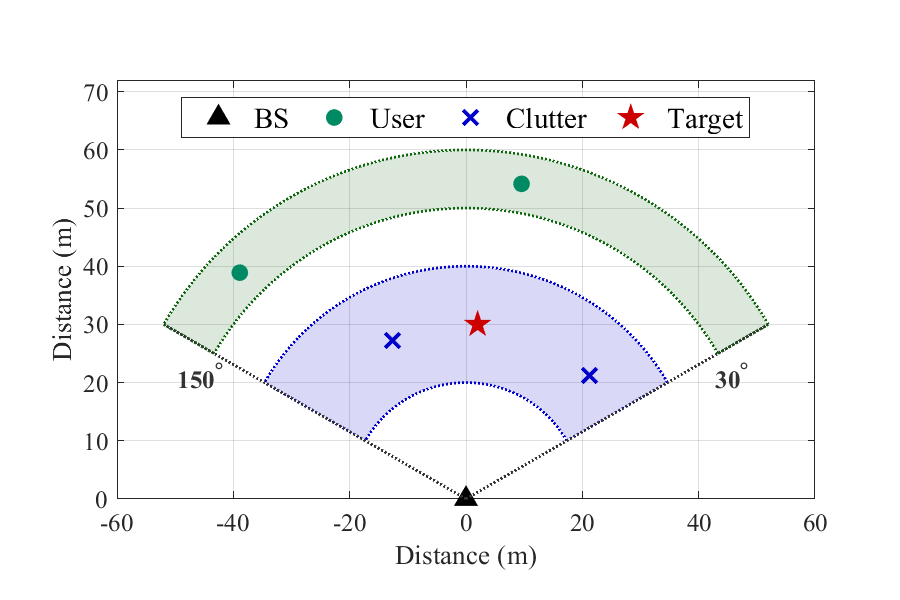}\\
  \caption{Locations of BS, users, target, and clutter scatterers.}
 \vspace{-0.5cm}
  \label{fig:setting}
\end{figure}

In this section, we present numerical simulations to validate the use of CRA-array-empowered ISAC systems using the proposed joint EM- and BB-domain optimization algorithm. Unless otherwise stated, the simulation parameters are set as follows: The BS, located at (0m, 0m), is equipped with separate transmit and receive arrays with $N=8$ antennas and operates at a carrier frequency of $f_{\mathrm{c}} = 28$ GHz with a total transmit power budget $P_{\mathrm{T}} = 60$ W.
Fig.~\ref{fig:setting} illustrates the considered scenario, where all entities are randomly distributed within an angular sector $[\pi/6,5\pi/6]$ relative to the BS. The $K=2$ users are randomly placed within an annular region centered at the BS, spanning radii from 50 m to 60 m. The target and clutter are located within the same closer annular region between 20 m and 40 m. The clutter environment comprises a total of $C=50$ scattering points, including two dominant clutter sources located at fixed positions $(45^{\circ}, 30\mathrm{m})$ and $(115^{\circ}, 30\mathrm{m})$. The remaining scatterers are modeled as weak distributed clutter points uniformly distributed over the angular domain.

The user communication channels are generated using the Quasi Deterministic Radio Channel Generator (QuaDRiGa), a recognized 3GPP TR 38.901 reference implementation. In particular, the channels are drawn from the ``3GPP\_3D\_UMi\_LOS'' scenario, with $L=5$ propagation paths per user. The polarization rotation matrices are assumed to be identity matrices \cite{Quadriga}. For the radar sensing channel, we adopt a standard line-of-sight (LoS) model incorporating path loss and depolarization effects. The path loss is given by
$10^{-\frac{C_{0}}{10}}\left(\frac{r_{i}}{D_{0}}\right)^{-\kappa_{i}}$,
where $r_{i}$ is the propagation distance, $\kappa_{i}=2.5$ denotes the attenuation factor, $C_{0}=30$ dB, $D_{0}=1$ m, and $i\in\{\mathrm{t},c\}$. Using the model in \cite{target polarization channel }, the depolarization channels (scattering matrices) for the target and clutter are modeled as zero-mean complex Gaussian random matrices, with their covariance matrix defined as:
\begin{equation}
\bm{\Sigma}_{\mathrm{t}}=\left[ \begin{array}{cccc} 0.1 & 0.06\varepsilon & 0.05\varepsilon & 0.04\varepsilon \\ 0.06\varepsilon^{\ast} & 0.3 & 0.03\varepsilon & 0.03\varepsilon \\ 0.05\varepsilon^{\ast} & 0.03\varepsilon^{\ast} & 0.1 & 0.03\varepsilon \\ 0.04\varepsilon^{\ast} & 0.03\varepsilon^{\ast} & 0.03\varepsilon^{\ast} & 1 \\ \end{array} \right],~\bm{\Sigma}_{c} = 0.2\mathbf{I}_{4},
\end{equation}
where $\varepsilon=1+j$.
The noise power at both users and BS is set to $\sigma_{k}^{2}= \sigma_{\mathrm{r}}^{2}=-80$ dBm, and the required communication SINR threshold is fixed at $\epsilon_{k}=5$ dB.

To illustrate the effectiveness of the proposed CRA array architecture, we evaluate its performance under two different reconfiguration resolutions. In the ideal high-resolution scenario, each CRA generates $P_{\mathrm{pat}} = 7$ distinct directional radiation patterns \cite{R. Murch 2022 pixel fig_reference_round} across $M = 180$ uniformly sampled azimuth angles. Combined with $P_{\mathrm{pol}} = 4$ polarization states \cite{H. Li 2024 polarization}, this configuration offers a total of $P = 28$ reconfigurable modes. We also investigate a lower-resolution scenario in which each CRA employs only $P_{\mathrm{pat}} = 3$ radiation patterns ($60^{\circ}$, $90^{\circ}$, $120^{\circ}$) and $P_{\mathrm{pol}} = 3$ polarization states (horizontal, vertical, $+45^{\circ}$), reducing the number of reconfigurable modes to $P = 9$.

\begin{figure}[t]
    \centering
    \subfigure[]{\includegraphics[width= 1.8 in]{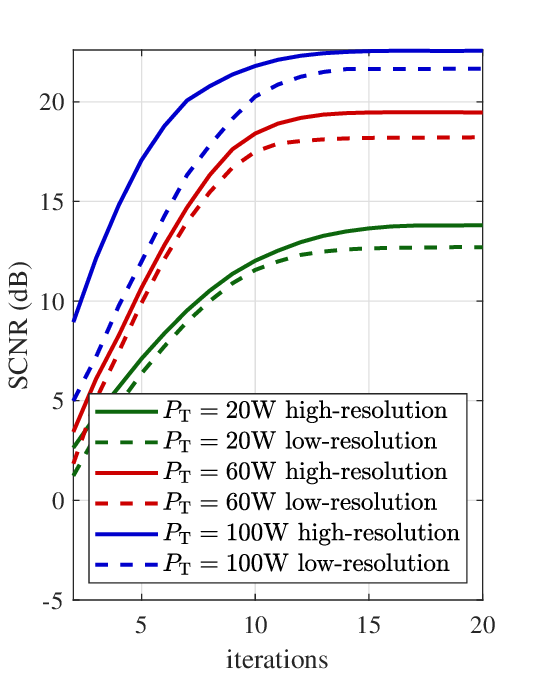}}%
    \subfigure[]{\includegraphics[width= 1.8 in]{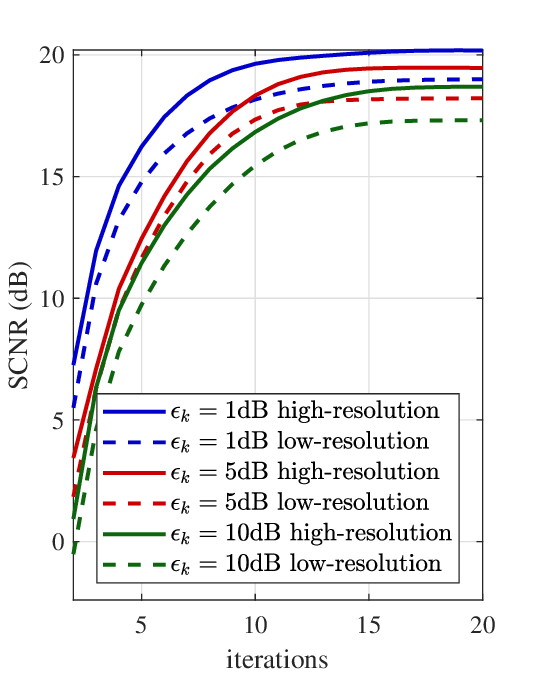}}
    \caption{Convergence performance.}
     \vspace{-0.5 cm}
    \label{fig:iteration}
\end{figure}

Fig.~\ref{fig:iteration}(a) presents the convergence behavior of the proposed joint EM- and BB-domain optimization algorithm for different transmit power budgets at the BS (20 W, 60 W, and 100 W), whereas Fig.~\ref{fig:iteration}(b) shows convergence under communication SINR thresholds ($\epsilon_{k}=1$ dB, 5 dB, 10 dB). The results clearly indicate that the proposed algorithm achieves rapid and stable convergence across all considered scenarios, underscoring its computational efficiency and practical applicability.

\begin{figure}[!t]
  \centering
  \includegraphics[width= 3.5 in]{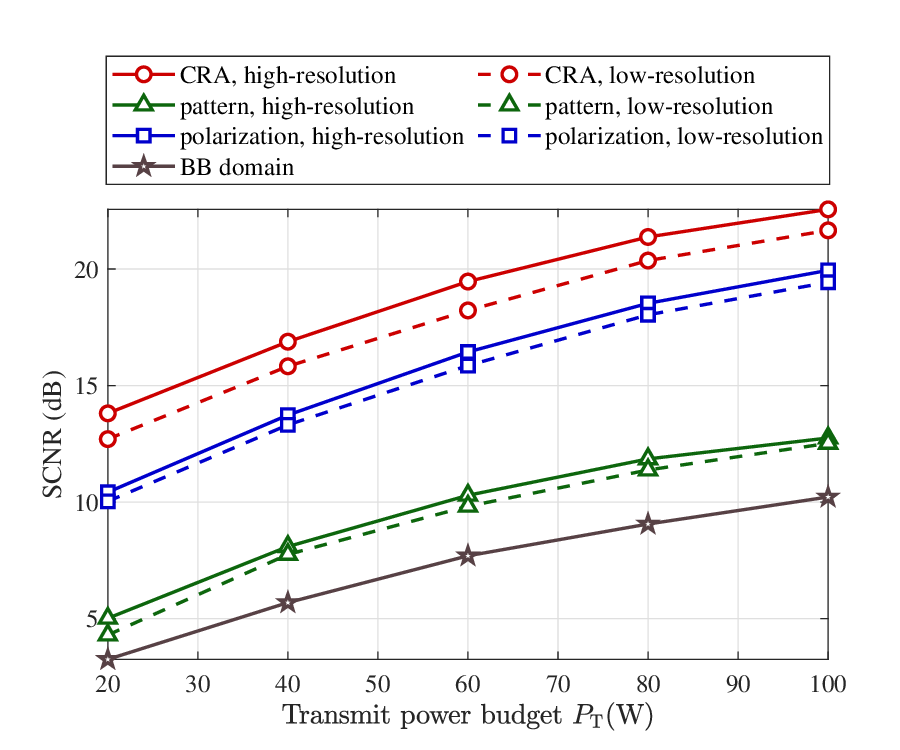}
  \caption{SCNR versus transmit power $P_\mathrm{T}$.}
  \label{fig:P}
   \vspace{-0.3 cm}
\end{figure}

Fig.~\ref{fig:P} illustrates radar SCNR performance as a function of transmit power for both the high- and low-resolution reconfigurable CRA scenarios, respectively labeled as \textbf{CRA, high-resolution} and \textbf{CRA, low-resolution}. To highlight the synergistic advantages of jointly reconfiguring the radiation patterns and polarization states, several baseline configurations are compared. Arrays utilizing only radiation-pattern reconfigurability  with fixed horizontal polarization are labeled as \textbf{pattern, high-resolution} and \textbf{pattern, low-resolution}, while those employing solely polarization reconfigurability with fixed boresight radiation pattern are labeled as \textbf{polarization, high-resolution} and \textbf{polarization, low-resolution}. Additionally, a baseline configuration employing fixed EM modes (boresight radiation pattern and horizontal polarization) with only BB-domain optimization is denoted as \textbf{BB domain}. The proposed CRA array approaches consistently outperform these baselines, demonstrating approximately 3 dB SCNR gain over polarization-only designs, around 9 dB improvement compared to pattern-only configurations, and up to 11 dB enhancement relative to pure BB-domain optimization. These results confirm that the proposed joint EM- and BB-domain optimization approach effectively improves radar sensing performance while meeting communication QoS constraints.

Moreover, the results in Fig.~\ref{fig:P} reveal that polarization reconfigurability offers more substantial performance gains than pattern reconfigurability. This superiority arises due to the inherent polarization differences between target and clutter scatterers, enabling effective interference suppression via polarization subspace alignment. Conversely, pattern reconfigurability primarily controls interference through angular sidelobe management, offering limited incremental benefits beyond traditional beamforming. Importantly, even low-resolution CRA implementations with fewer discrete polarization and pattern modes achieve near-optimal performance, highlighting the viability of low-complexity CRA implementations.

\begin{figure}[!t]
    \centering
    \vspace{0.2 cm}
   \includegraphics[width= 3.5 in]{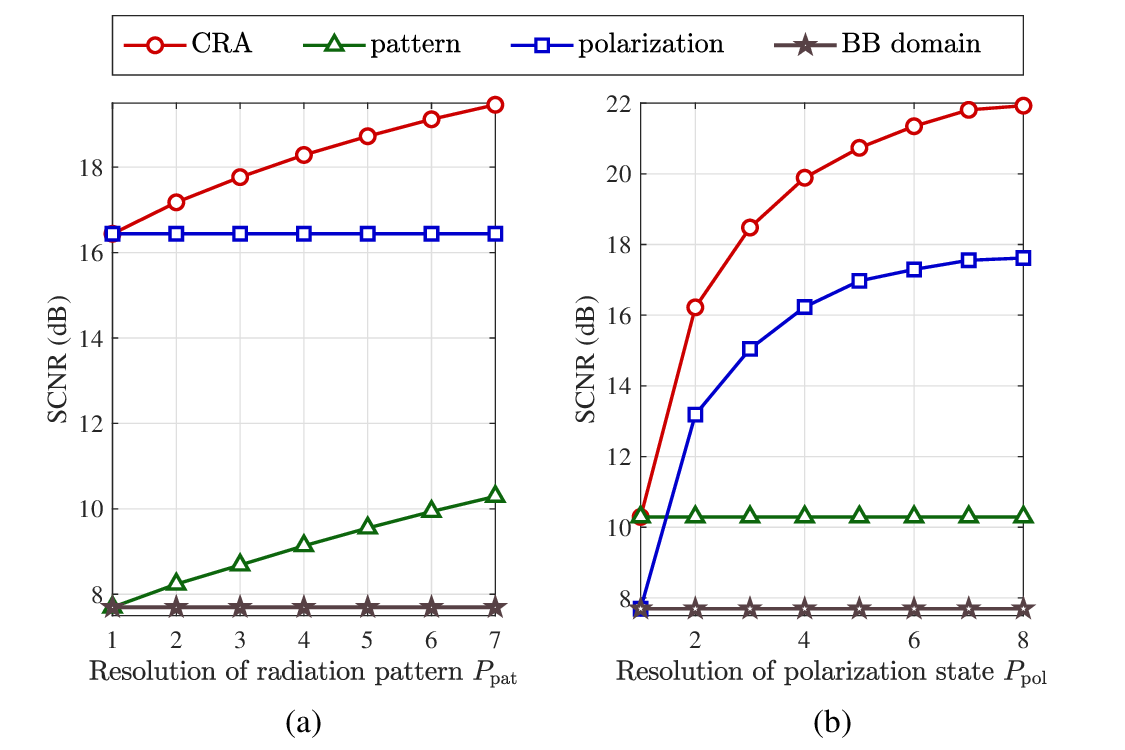}%
    \caption{SCNR versus the resolution of radiation patterns and polarization states.}
     \vspace{-0.4 cm}
    \label{fig:resolution}
\end{figure}

Fig.~\ref{fig:resolution} examines the impact of radiation pattern and polarization resolutions on radar SCNR performance. Fig.~\ref{fig:resolution}(a) shows that increasing pattern resolution provides SCNR improvements of around 3 dB for the CRA design and approximately 2 dB for the pattern-only design. In contrast, Fig.~\ref{fig:resolution}(b) indicates that higher polarization resolution delivers more significant gains, approximately 12 dB for the CRA configuration and around 10 dB for the polarization-only scheme. These findings reinforce the greater importance of polarization over pattern reconfigurability, consistent with earlier observations in Fig.~\ref{fig:P}.

\begin{figure}[!t]
  \centering
  \includegraphics[width= 3.5 in]{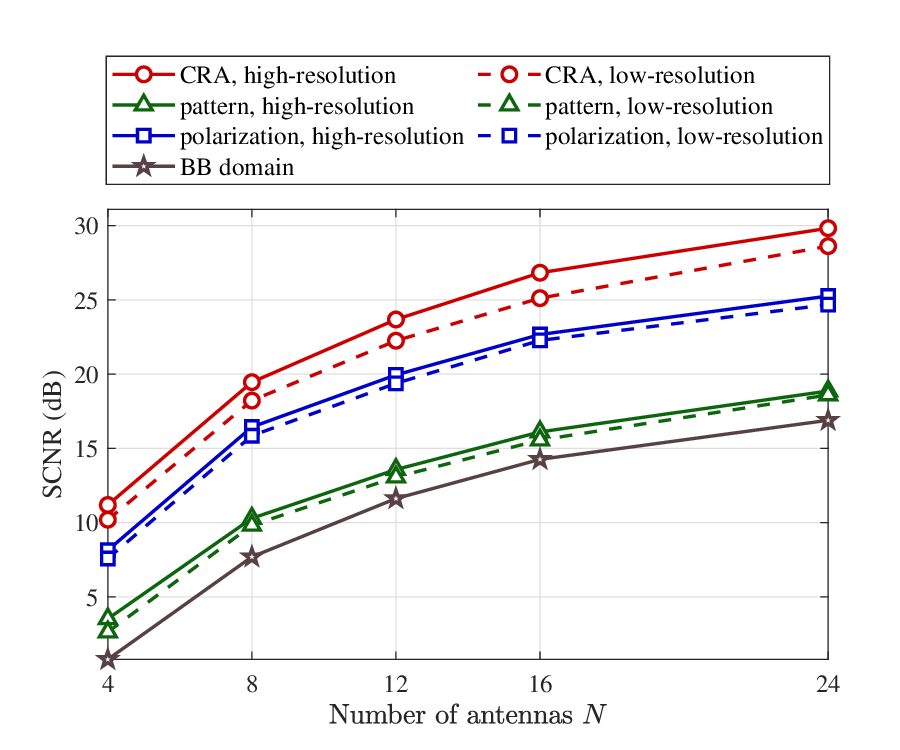}
  \caption{SCNR versus the number of antennas  $N$.}
  \label{fig:N}
   \vspace{-0.4 cm}
\end{figure}

Fig.~\ref{fig:N} presents the SCNR performance as a function of the number of transmit/receive antennas, demonstrating the inherent advantage of reconfigurability in terms of array size. In particular, the CRA array design achieves performance comparable to that of conventional antenna arrays employing only BB-domain optimization with up to 75\% fewer antennas. These results clearly highlight the hardware efficiency and size-reduction benefits of CRA arrays, reinforcing their potential to support energy-efficient next-generation wireless systems without compromising overall performance.

\begin{figure}[t]
    \centering
   \includegraphics[width= 3.5 in]{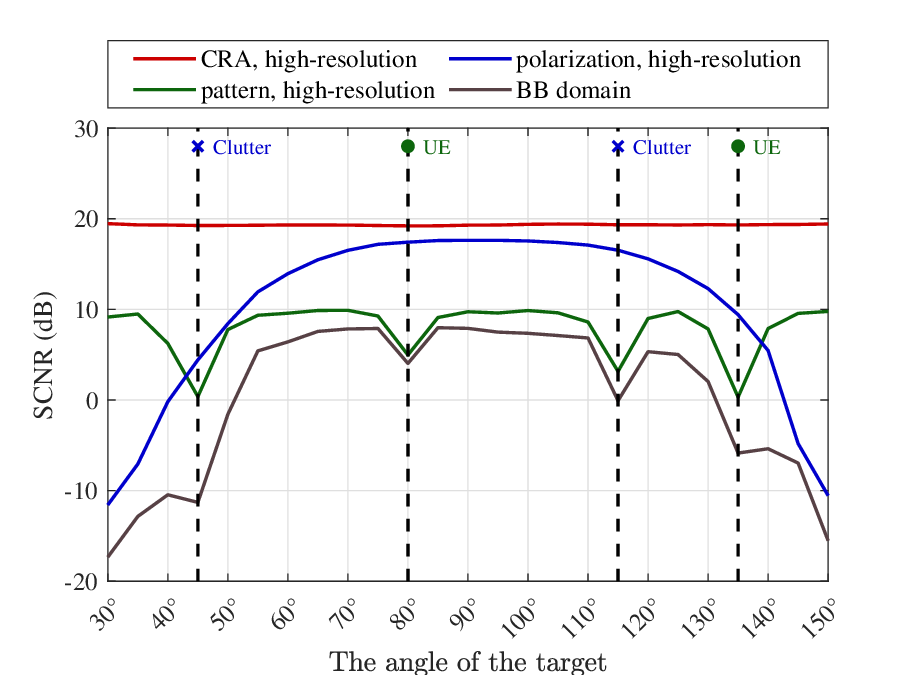}%
    \caption{SCNR versus target angle.}
     \vspace{-0.3 cm}
    \label{fig:angle}
\end{figure}

Fig. \ref{fig:angle} presents radar sensing performance as a function of the target angle, with fixed positions for users and two dominant clutter objects. Two users are placed at polar coordinates $(80^{\circ}, 55\mathrm{m})$ and $(135^{\circ}, 55 \mathrm{m})$, while two clutter sources are positioned at $(45^{\circ}, 30\mathrm{m})$ and $(115^{\circ}, 30 \mathrm{m})$. The proposed ``{CRA, high-resolution}'' approach consistently achieves robust radar performance across a wide angular range. The ``{polarization, high-resolution}'' implementation maintains satisfactory sensing performance at angles coincident with clutter scatterers or users, owing to its polarization flexibility, but experiences performance degradation at more extreme angles due to the absence of pattern reconfigurability. In contrast, the performance of the ``{pattern, high-resolution}'' and ``{BB domain}'' approaches degrade considerably when the target angle overlaps with the sources of interference. These results confirm the effectiveness of polarization diversity in resolving targets that are near strong clutter scattering.

\begin{figure}[t]
    \centering
   \includegraphics[width= 3.5 in]{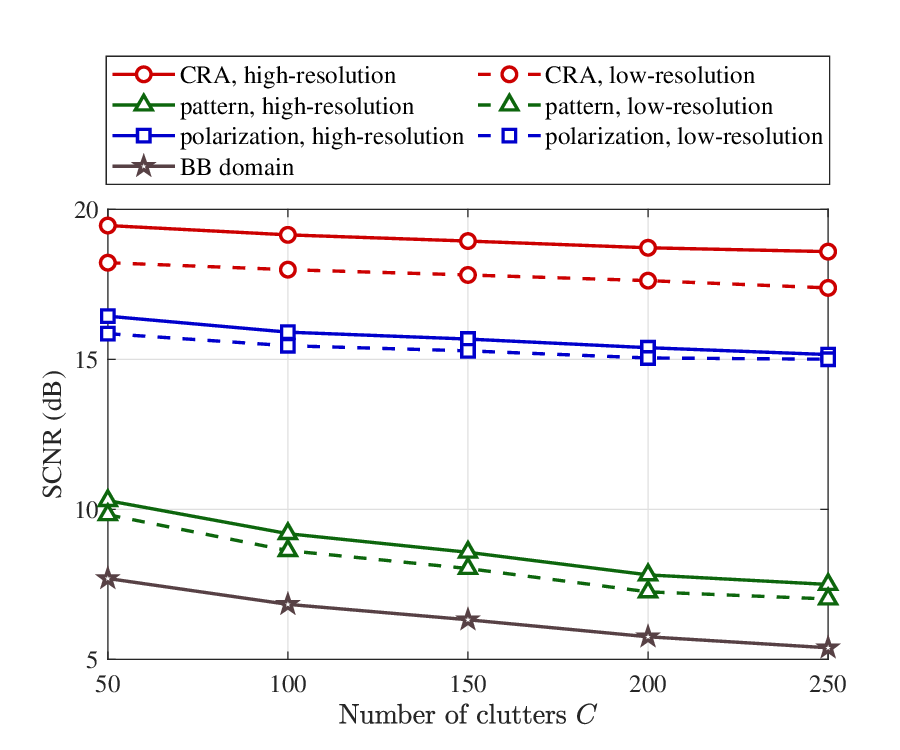}%
    \caption{SCNR versus the number of clutter scatterers.}
     \vspace{-0.4 cm}
    \label{fig:clutter}
\end{figure}

Fig. \ref{fig:clutter} illustrates radar sensing performance versus the number of clutter scattering points. As the number of scatterers increases, the sensing SCNR of pure BB-domain optimization drops by about 3 dB. This degradation stems from the reduced number of available spatial DoFs for clutter suppression and results in stronger residual clutter at the radar output. Conversely, the degradation of the CRA design is more gradual (less than 0.5 dB) as the clutter density grows, thanks to the additional EM-domain DoFs provided by radiation pattern and polarization reconfigurability. Thus, even in dense clutter conditions, the CRA design better preserves target–clutter separability, leading to improved robustness.

\begin{figure}[t]
    \centering
   \includegraphics[width= 3.5 in]{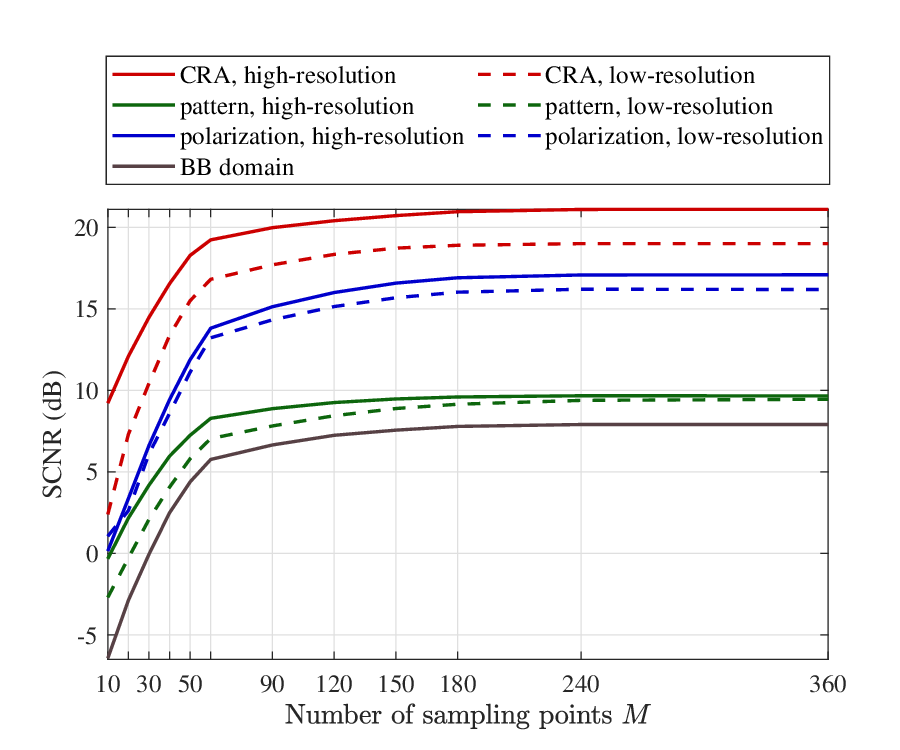}%
    \caption{SCNR versus number of angular sampling points $M$.}
     \vspace{-0.5 cm}
    \label{fig:M}
\end{figure}

In Fig. \ref{fig:M}, we investigate how the number of angular sampling points $M$ affects the radar SCNR. It is evident that the SCNR will increase with $M$, since a finer angular grid provides a more accurate discretization of continuous radiation patterns, thus improving the matching precision between actual radiation gains and object angles. Nevertheless, the incremental benefit gradually becomes negligible as
$M$ increases beyond a certain point around $M=180$, with only 2 dB of loss at $M=60$, indicating limited value in excessively dense angular sampling. The choice of
$M=180$ adopted in earlier simulations achieves a favorable balance between performance improvement and computational complexity.

\begin{figure}[t]
    \centering
   \includegraphics[width= 3.5 in]{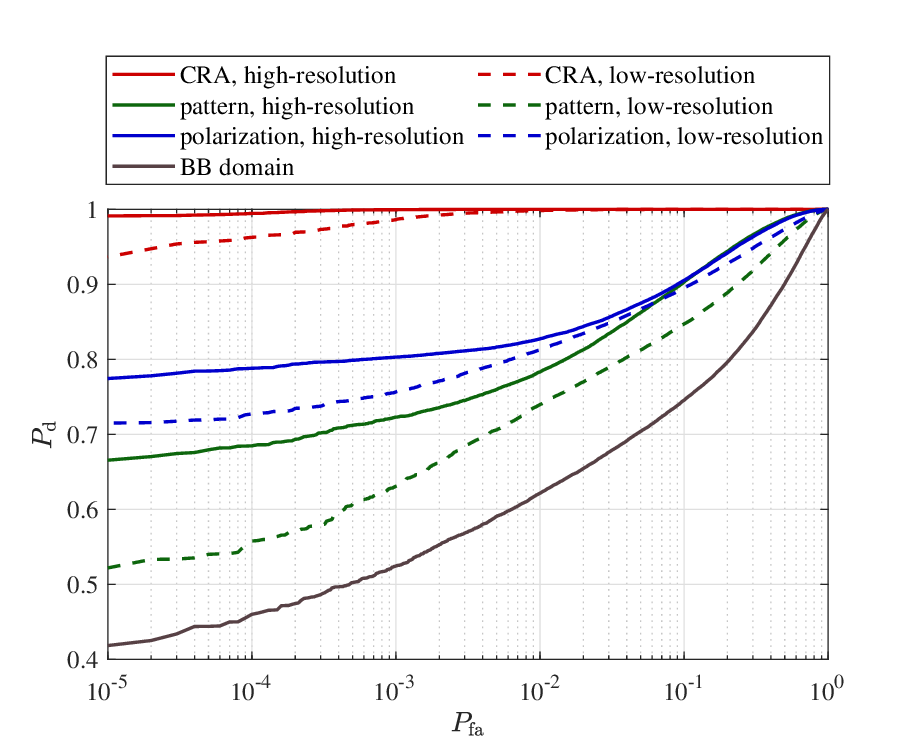}%
    \caption{ROC curve for various algorithms.}
     \vspace{-0.4 cm}
    \label{fig:ROC}
\end{figure}

Finally, Fig.~\ref{fig:ROC} presents the receiver operating characteristic (ROC) curves, illustrating target detection probability ($P_{\mathrm{d}}$) against false alarm probability ($P_{\mathrm{fa}}$). The results clearly show that the proposed joint EM- and BB-domain optimization algorithm significantly outperforms all benchmark schemes. Specifically, for false alarm probabilities below $10^{-3}$, the proposed approach achieves over 50\% improvement in detection probability compared to conventional BB-domain beamforming. Due to the limited available spatial resolution with only $N=8$ receive antennas, the BB-domain beamforming design fails to adequately discriminate closely spaced targets, users, and clutter. In contrast, the proposed CRA array, leveraging both radiation-pattern and polarization reconfigurability, consistently achieves detection probabilities approaching 99\% at $P_{\mathrm{fa}} = 10^{-3}$, even for the low-resolution implementation.

\section{Conclusions}
In this paper, we proposed a novel ISAC system architecture empowered by CRAs. By simultaneously enabling dynamic reconfiguration of radiation patterns and polarization states, CRAs significantly expand the number of available EM DoFs. We developed a comprehensive compound channel model integrating virtual angular-domain characteristics, spatial propagation, and depolarization effects. Based on this model, we formulated a joint EM- and BB-domain optimization framework to maximize radar sensing performance subject to communication QoS, transmit power, and SINR constraints. To tackle the resulting MINLP problem, we proposed a decomposition-based iterative algorithm leveraging FP, MM, SOCP, and penalty methods.
Extensive simulations demonstrated that the proposed CRA-array architecture effectively enhances radar sensing accuracy and clutter suppression compared to conventional beamforming approaches. The results highlighted the critical importance of polarization reconfigurability and confirmed that even practical low-resolution CRA implementations offer near-optimal performance. Thus, this work underscores the significant advantages of exploiting additional EM-domain DoFs, demonstrating EM reconfigurability to be a promising technology for robust and efficient next-generation ISAC systems. Our research bridges theoretical ISAC design with practical antenna implementation constraints, facilitating the development of resilient and high-performance 6G wireless networks.

\end{document}